\documentclass[seceq,paper,dvipdfmx,twocolumn]{jpsj3}
\usepackage{color,bm,amsmath,amssymb,graphicx,url,float}

\renewcommand{\H}{{\mathcal H}}

\newcommand{\bS}{{\bm{S}}}

\newcommand{\bk}{{\bm{k}}}

\newcommand{\bd}{{\bm{d}}}
\newcommand{\bt}{{\bm{t}}}

\newcommand{\ba}{{\bm{a}}}
\newcommand{\bA}{{\bm{A}}}
\newcommand{\bB}{{\bm{B}}}
\newcommand{\bLambda}{{\bm{\Lambda}}}
\newcommand{\bPi}{{\bm{\Pi}}}
\newcommand{\bDelta}{{\bm{\Delta}}}
\newcommand{\rd}{{\rm d}}

\newcommand{\bu}{{\bm u}}
\newcommand{\bv}{{\bm v}}
\newcommand{\balpha}{{\bm \alpha}}
\newcommand{\bU}{{\bm U}}
\newcommand{\bV}{{\bm V}}

\newcommand{\bM}{{\bm M}}
\newcommand{\bh}{{\bm h}}

\title{Edge Magnon Excitation in Spin Dimer Systems}

\author{Ryo Sakaguchi and Masashige Matsumoto\thanks{E-mail address: matsumoto.masashige@shizuoka.ac.jp}}

\inst{Department of Physics, Shizuoka University, Shizuoka 422-8529, Japan}

\recdate{May 20, 2016}

\abst{
Magnetic excitation in a spin dimer system on a bilayer honeycomb lattice is investigated in the presence of a zigzag edge,
where disordered and ordered phases can be controlled by a quantum phase transition.
In analogy with the case of graphene with a zigzag edge, a flat edge magnon mode appears in the disordered phase.
In an ordered phase, a finite magnetic moment generates a mean-field potential to the magnon.
Since the potential is nonuniform on the edge and bulk sites,
it affects the excitation, and the dispersion of the edge mode deviates from the flat shape.
We investigate how the edge magnon mode evolves when the phase changes through the quantum phase transition
and discuss the similarities to ordered spin systems on a monolayer honeycomb lattice.
}

\begin{document}

\maketitle

\section{Introduction}

It is well known that a flat band appears in a graphene in the presence of a zigzag edge.
\cite{Fujita-1996}
It was pointed out by Fujita and coworkers
that the wavefunction of the $\pi$ electron of the flat band is localized at the edge
while showing an exponential decrease in its amplitude away from the edge.
\cite{Fujita-1996,Nakada-1996}
In contrast, no localized edge state appears in the case of an armchair edge.
This work has attracted much interest and the investigation of the edge state in graphene has been carried out.
The localized edge state was experimentally observed in the vicinity of the zigzag edge
by scanning tunneling microscope measurements.
\cite{Kobayashi-2005,Niimi-2006}
Theoretically, it was pointed out that the existence of the flat edge band
is related to the topological nature of the bulk system.
\cite{Ryu-2002,Hatsugai-2009,Yao-2009,Delplace-2011}
The concept of the edge state was also applied to insulating systems such as quantum magnets.
The edge magnetic excitation was also investigated from the topological viewpoint.
\cite{Zhang-2013,Mook-2014,Chisnell-2015}

As in the case of graphene, a localized magnetic excitation can be expected
in quantum spin systems on a honeycomb lattice with a zigzag edge.
For a monolayer antiferromagnetic (AF) system, You et al. examined spin-wave excitation
on the basis of Holstein--Primakoff theory and found a dispersive localized magnon near the zigzag edge.
\cite{You-2008}
The Hamiltonian for graphene is described by a fermion for the $\pi$ electron.
The electron moves on the honeycomb lattice by a hopping process.
The local energy of the $\pi$ electron is assumed to be uniform in the case of graphene.
In contrast, a magnetic excitation moves via the pair creation and annihilation process of bosons
for the AF spin system.
In addition, the local energy for the excited state is nonuniform.
This is owing to the fact that the number of nearest-neighbor sites is different between the edge and bulk sites.
This makes the mean-field potential different on these sites in the ordered phase.
Thus, there are two differences between the graphene and the AF spin system.

In this paper, we focus on a model that can connect the two cases.
A spin dimer system is a good example for studying the edge magnetic excitation for the following reason.
In spin dimer systems, the magnetic excitation moves via both the hopping and pair creation and annihilation processes.
Another characteristic point is that the disordered and ordered phases can be controlled by a quantum phase transition
driven by applying a magnetic field or pressure.
The mean-field potential is uniform in the disordered phase, while it is nonuniform in the ordered phase.
Thus, to study the spin dimer systems gives variety to the investigation of the edge mode
and helps us understand how the edge mode emerges and evolves in quantum spin systems
when the phase changes through the quantum phase transition.
For this purpose, we consider a bilayer honeycomb lattice.
The interlayer interaction is assumed to be strong and spin dimers are formed perpendicular to the layer.
The dimers interact with each other through interdimer interactions along the honeycomb lattice.
Note that this type of spin system on an isolated bilayer honeycomb lattice was synthesized recently
in Bi$_3$Mn$_4$O$_{12}$(NO$_3$).
\cite{Smirnova-2009}
This system is an $S=3/2$ spin dimer system showing no magnetic phase transition down to low temperatures.
\cite{Smirnova-2009,Okubo-2010,Onishi-2012}
To identify the exchange interaction parameters,
the stability of the ordered phase has been investigated theoretically in connection with the phase diagram.
\cite{Kandpal-2011,Ganesh-2011,Oitmaa-2012}
The time has come to study the edge magnon in spin dimer systems on a bilayer honeycomb lattice.

In our study, we consider an $S=1/2$ spin dimer system on a bilayer honeycomb lattice with a zigzag edge.
To concentrate on the edge magnetic excitation at zero temperature,
we apply a bond operator formulation on the basis of the dimer mean-field theory developed by Shiina et al.
\cite{Shiina-2003}
This formulation is equivalent to the formulation applied to interacting spin dimer systems.
\cite{Sommer-2001,Matsumoto-2002,Matsumoto-2004}
We report that an edge magnon mode with a completely flat dispersion relation appears in the disordered phase.
The wavefunction of the edge mode is examined analytically.
In an ordered phase, the flat mode changes its shape to a dispersive one.
We compare our results with those for graphene and monolayer spin systems
and discuss the similarities and differences between them.

This paper is organized as follows.
In Sect. 2, we briefly summarize a bulk property of the spin dimer system on an infinite bilayer honeycomb lattice.
In Sect. 3, we give a formulation for the magnetic excitation in the presence of the zigzag edge.
In Sect. 4, we study the disordered phase.
A edge magnon mode with a completely flat dispersion relation is reported as a result of an analytic investigation.
In Sect. 5, it is reported how the edge magnon mode evolves in the ordered phase.
To understand the dispersive edge mode in the ordered phase,
spin systems on a monolayer honeycomb lattice are studied in Sect. 6.
Section 7 gives summary and discussion.
Details of the formulation are given in the Appendices.

\section{Spin Dimers on Infinite Honeycomb Lattice}

\begin{figure}[t]
\begin{center}
\includegraphics[width=8cm,clip]{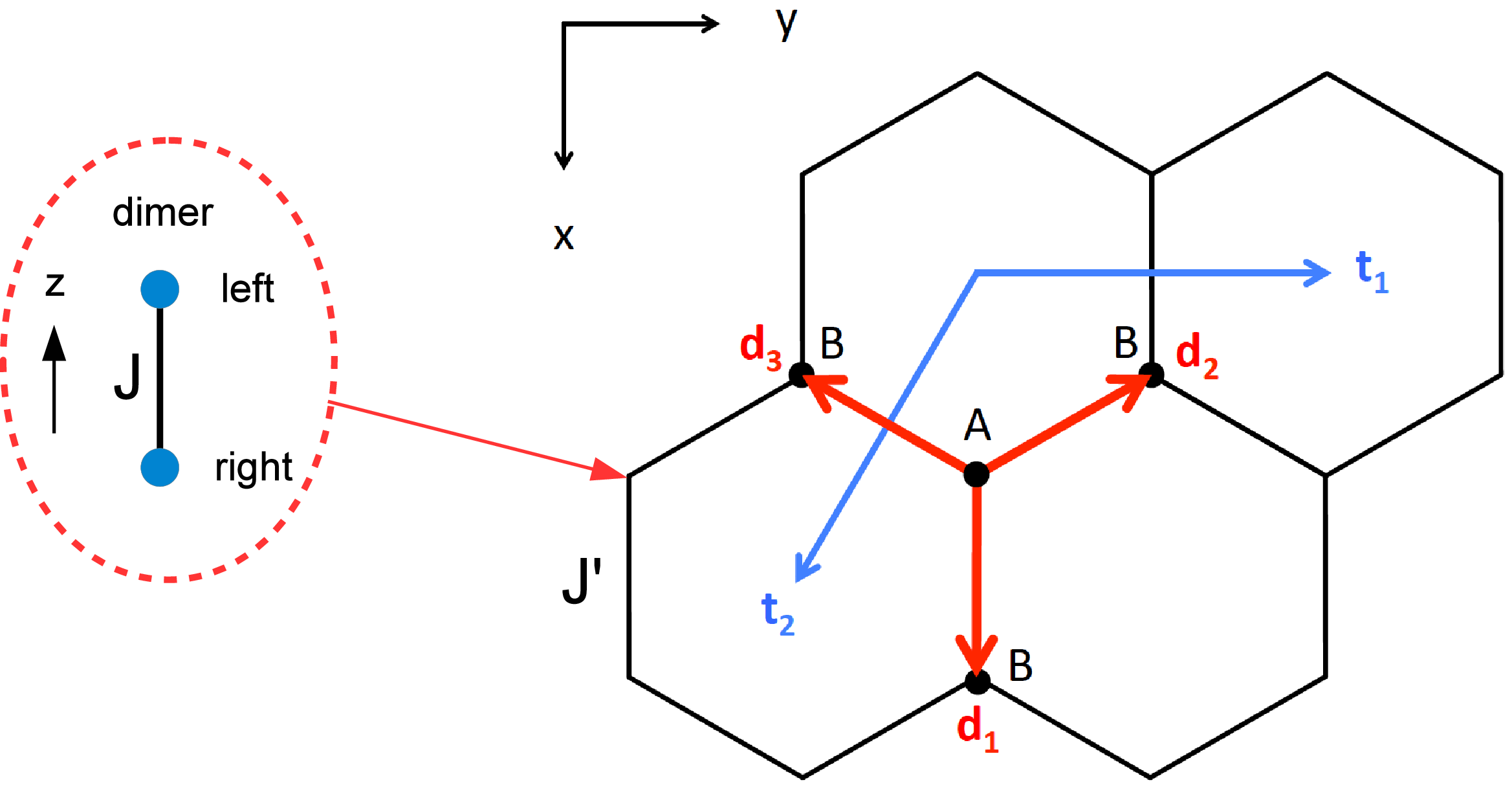}
\end{center}
\begin{center}
\vspace*{5mm}
\includegraphics[width=3.5cm,clip]{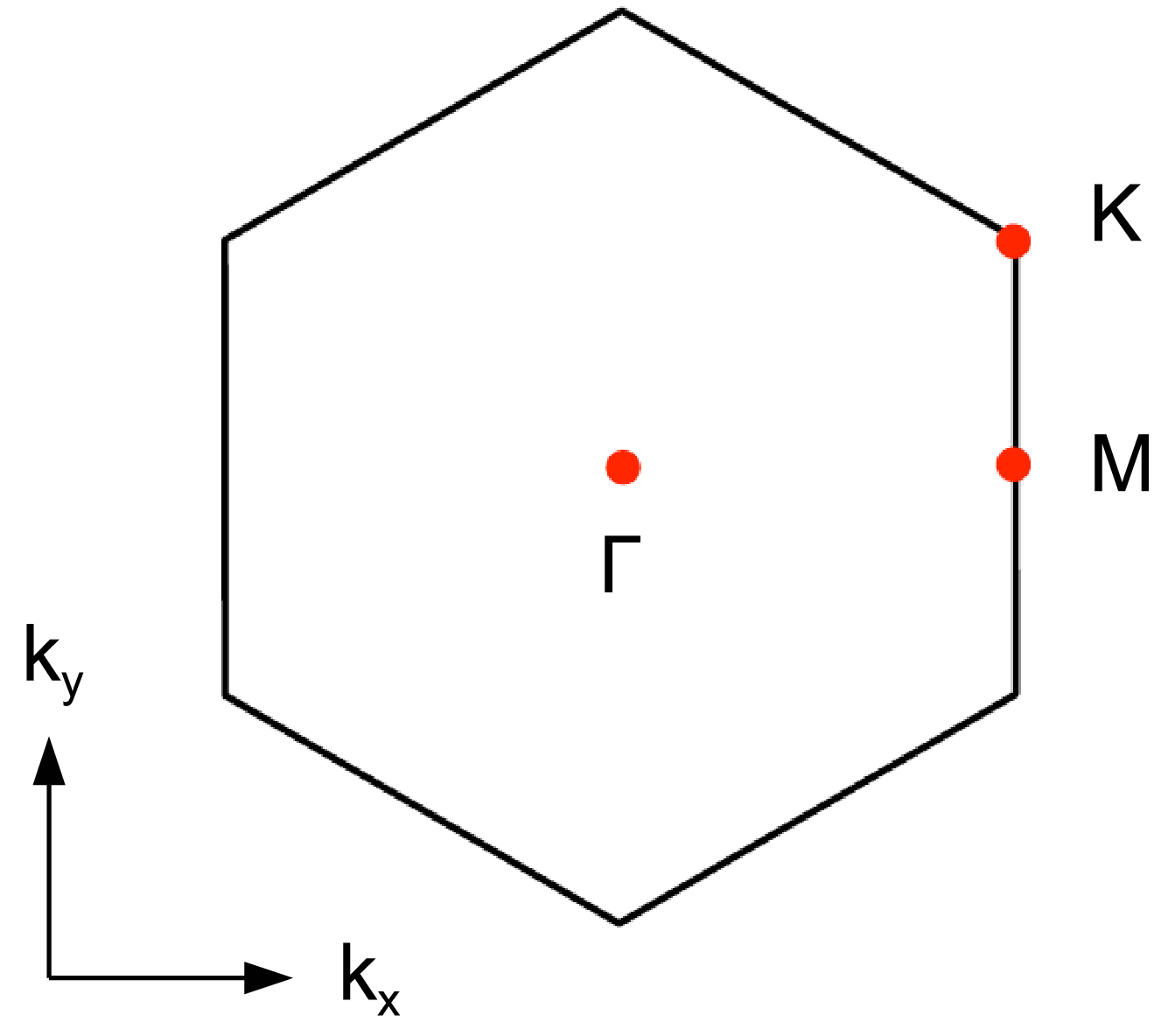}
\end{center}
\caption{
(Color online)
Schematic of an infinite bilayer honeycomb lattice.
A dimer is formed by the left- and right-side spins, which are aligned perpendicular to the layer.
They are coupled by the intradimer interaction $J$.
The dimers are aligned along the $z$-direction and are located at every apex of the hexagons on the honeycomb lattice.
The interdimer interaction $J'$ connects the same side (left or right) of dimers along the honeycomb lattice.
$\bt_1$ and $\bt_2$ are translational vectors defined by
$\bt_1 = (0,1)$ and $\bt_2 = (\frac{\sqrt{3}}{2},- \frac{1}{2})$ in the lattice constant unit.
The honeycomb lattice is divided into two sublattices, A and B.
They are connected by the vectors
$\bd_1 = (\frac{1}{\sqrt{3}},0)$, $\bd_2 = (-\frac{1}{2\sqrt{3}},\frac{1}{2})$,
$\bd_3 = (-\frac{1}{2\sqrt{3}},- \frac{1}{2})$.
The lower panel represents the reciprocal lattice.
The $\Gamma$, M, and K points are defined by
$\Gamma=(0,0)$, ${\rm M}=(\frac{2\pi}{\sqrt{3}},0)$, and ${\rm K}=(\frac{2\pi}{\sqrt{3}},\frac{2\pi}{3})$
in the reciprocal lattice unit.
}
\label{fig:honeycomb}
\end{figure}

Before examining the edge magnon excitation, we briefly study a bulk property of spin dimers
on an infinite bilayer honeycomb lattice as shown in Fig. \ref{fig:honeycomb}.
The two layers are connected by the intradimer interaction $J$,
while the interdimer interaction $J'$ is along the honeycomb lattice.
The model Hamiltonian is expressed as
\begin{align}
&\H = \H_{\rm intra} + \H_{\rm inter}, \cr
&\H_{\rm intra}
= \sum_i \left[ J \bS_{i{\rm l}} \cdot \bS_{i{\rm r}} - \bm{h}\cdot( \bS_{i{\rm l}} + \bS_{i{\rm r}} ) \right], \cr
&\H_{\rm intrer} = \sum_{\langle ij\rangle} J'
\left( \bS_{i{\rm l}} \cdot \bS_{j{\rm l}} + \bS_{i{\rm r}} \cdot \bS_{j{\rm r}} \right).
\label{eqn:H-dimer}
\end{align}
Here, $\bS_{i{\rm l}}$ and $\bS_{i{\rm r}}$ are $S=1/2$ spin operators
on the left and right sides of a dimer at the $i$th site, respectively.
The summation $\sum_{\langle ij\rangle}$ is taken over the spin pairs at the nearest-neighbor sites.
$\H_{\rm intra}$ and $\H_{\rm inter}$ represent the intradimer and interdimer interaction parts of the Hamiltonian, respectively.
$\bm{h}=g\mu_{\rm B}\bm{H}$ represents the effective magnetic field,
where $\mu_{\rm B}$ and $\bm{H}$ are the Bohr magneton and external magnetic field, respectively.
Since the Hamiltonian has rotational symmetry, we take the $z$-axis along the external field.

\begin{figure}[t]
\begin{center}
\includegraphics[width=7cm,clip]{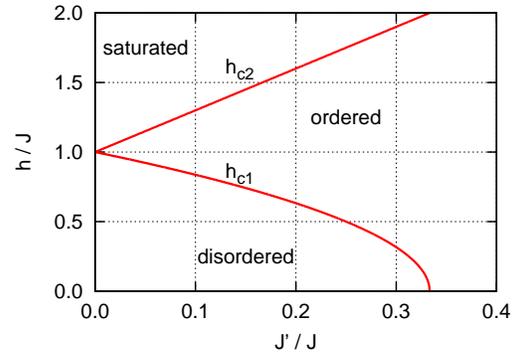}
\end{center}
\caption{
(Color online)
$J'/J$ dependence of the critical fields $h_{\rm c1}$ and $h_{\rm c2}$.
Their expressions are respectively given by Eqs. (\ref{eqn:hc1}) and (\ref{eqn:hc2}).
}
\label{fig:phase}
\end{figure}

\begin{figure}[t]
\begin{center}
\includegraphics[width=8cm,clip]{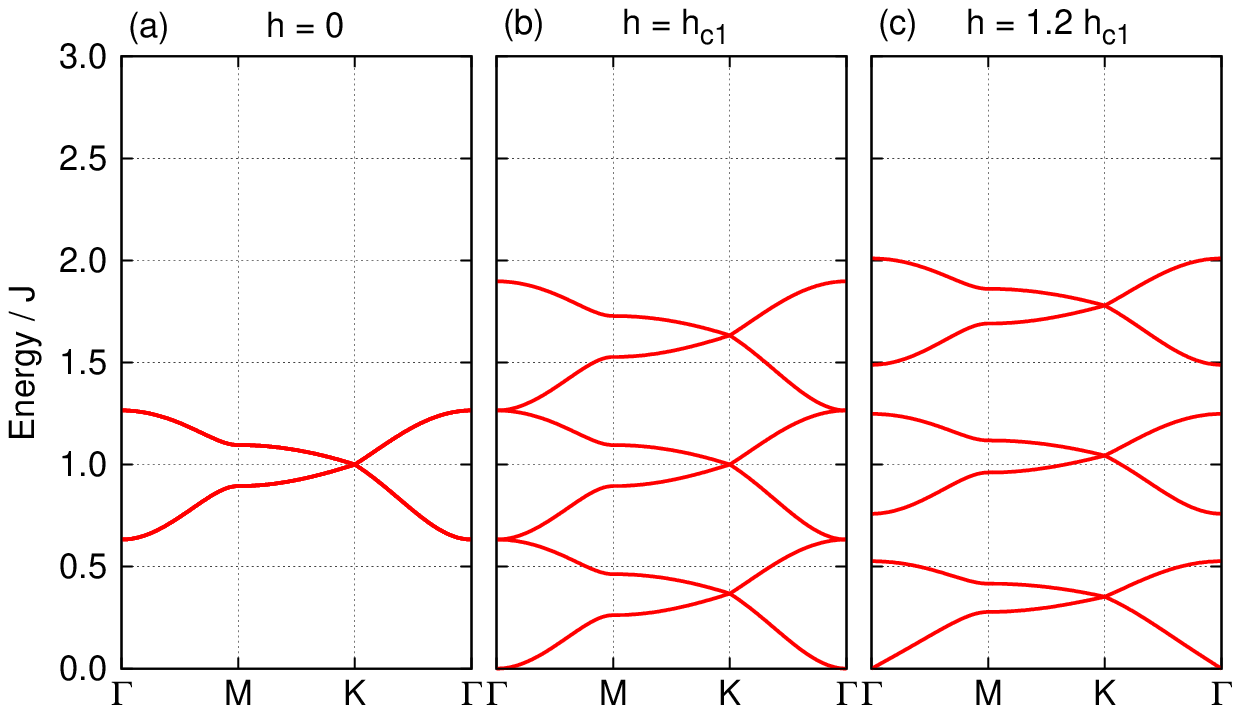}
\end{center}
\begin{center}
\includegraphics[width=8cm,clip]{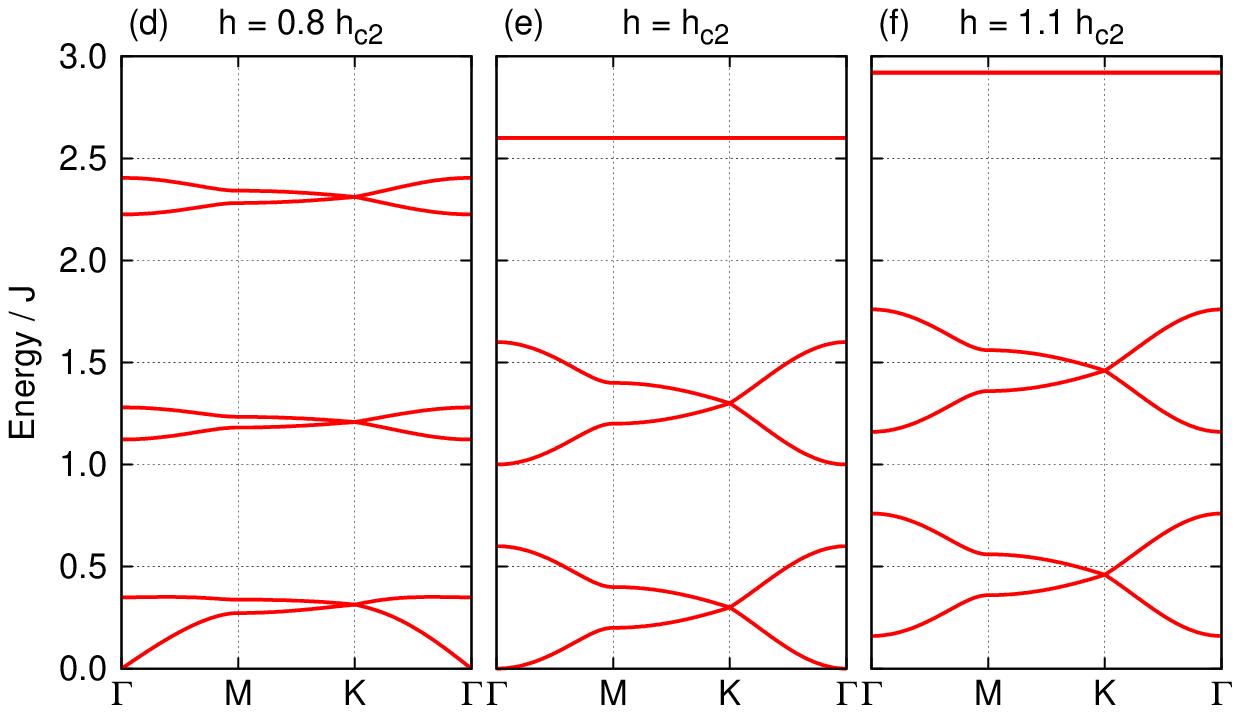}
\end{center}
\caption{
(Color online)
Magnon dispersion relation under various magnetic fields for $J'/J=0.2$.
(a) For $h=0$.
(b) For $h=h_{\rm c1}$.
(c) For $h=1.2h_{\rm c1}$.
(d) For $h=0.8h_{\rm c2}$.
(e) For $h=h_{\rm c2}$.
(f) For $h=1.1h_{\rm c2}$.
The $\Gamma$, M, and K points in the reciprocal lattice are given in the caption of Fig. \ref{fig:honeycomb}.
Equation (\ref{eqn:H-final-2}) is used in the plot for $h\le h_{\rm c1}$.
}
\label{fig:magnon-bulk-h}
\end{figure}

For the infinite lattice, magnetic excitations are expressed analytically
in the disordered phase on the basis of bond operator theory.
Details are given in Appendix \ref{appendix-bulk}.
For small $J'$ and $h$, a singlet ground state is stabilized.
Increasing them induces a finite magnetic moment and leads to an AF ordered phase.
Figure \ref{fig:phase} shows the phase diagram obtained by dimer mean-field theory.
$h_{\rm c1}$ is the lower critical field, above which the ordered phase appears,
while $h_{\rm c2}$ is the higher critical field, above which the moment saturates.

First, we study the case of field-induced order.
We show the magnon dispersion relation in Fig. \ref{fig:magnon-bulk-h} for various values of $h$.
At $h=0$, there is a finite excitation gap at the $\Gamma$ point as shown in Fig. \ref{fig:magnon-bulk-h}(a).
There are two modes, reflecting the two dimer sites in the unit cell.
They have the same energy at the K point.
Each mode is threefold degenerate, reflecting the triplet excitation.
The degeneracy is lifted by a finite field, and the excitation mode splits into three branches.
At the lower critical field ($h=h_{\rm c1}$), the lowest mode becomes soft,
showing a quadratic dispersion relation at the $\Gamma$ point [see Fig. \ref{fig:magnon-bulk-h}(b)].
In the ordered phase, a finite AF moment appears and the lowest mode shows a linear dispersion relation
[see Figs. \ref{fig:magnon-bulk-h}(c) and \ref{fig:magnon-bulk-h}(d)].
In the ordered phase, we applied an extended spin-wave theory,
\cite{Shiina-2003}
which is also used to study the edge magnon in the subsequent sections.
At the higher critical field ($h=h_{\rm c2}$),
the lowest mode shows a quadratic dispersion relation again [see Fig. \ref{fig:magnon-bulk-h}(e)].
In the saturated phase ($h>h_{\rm c2}$), the excitation gap opens up again [see Fig. \ref{fig:magnon-bulk-h}(f)].
In Figs. \ref{fig:magnon-bulk-h}(e) and \ref{fig:magnon-bulk-h}(f),
the highest modes are twofold degenerate and become dispersionless.
Since the unit cell is taken, as in the ordered phase,
there are two dimer sites in a unit cell and this leads to twofold degeneracy.
Note that the flat modes are not edge states.
For $h\ge h_{\rm c2}$, the local ground state is the $|\uparrow\uparrow\rangle$ triplet state of a dimer,
while the highest mode is the $|\downarrow\downarrow\rangle$ triplet state.
Within the harmonic approximation in bond operator theory,
the excited $|\downarrow\downarrow\rangle$ state does not propagate under the $|\uparrow\uparrow\rangle$ ground state
and it becomes dispersionless.

\begin{figure}[t]
\begin{center}
\includegraphics[width=8cm,clip]{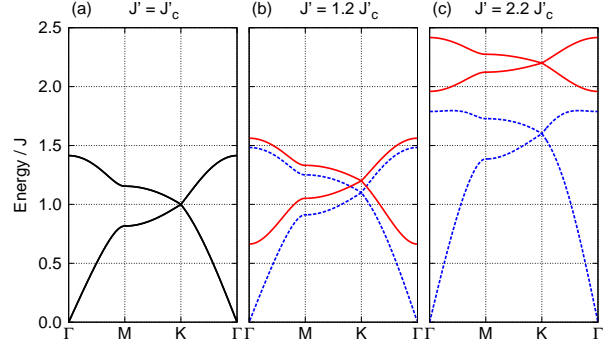}
\end{center}
\caption{
(Color online)
Magnon dispersion relation for various values of the interdimer interaction under $h=0$.
(a) For $J'=J'_{\rm c}$.
The critical interdimer interaction is defined as $J'_{\rm c}=\frac{1}{3}J$.
The excitation is threefold degenerate for $J'\le J'_{\rm c}$.
(b) For $J'=1.2J'_{\rm c}$.
(c) For $J'=2.2J'_{\rm c}$.
The excitation splits into a twofold T-mode (blue dashed line) and a single L-mode (red solid line) for $J'> J'_{\rm c}$.
Equation (\ref{eqn:E-order}) is used for the plot.
}
\label{fig:magnon-bulk-jp}
\end{figure}

The ordered phase is also stabilized even at $h=0$.
The magnon band becomes wide when $J'$ is increased.
At the critical interdimer interaction ($J'=J'_{\rm c}$),
the lowest mode becomes soft, showing a linear dispersion relation at the $\Gamma$ point
while keeping the threefold degeneracy [see Fig. \ref{fig:magnon-bulk-jp}(a)].
For $J'>J_{\rm c}'$, an AF moment appears and the magnon modes split into a twofold degenerate transverse mode (T-mode)
and a longitudinal mode (L-mode or Higgs amplitude mode) as shown in Fig. \ref{fig:magnon-bulk-jp}(b).
The splitting of the two modes becomes large with increasing $J'$ [see Fig. \ref{fig:magnon-bulk-jp}(c)].
In the ordered phase under $h=0$, the formulation becomes simple
and the dispersion relations for the L- and T-modes are given analytically in Appendix \ref{appendix-bulk}.

\section{Formulation for Edge Magnon Mode}

\subsection{Model Hamiltonian}

\begin{figure}[t]
\begin{center}
\includegraphics[width=7cm,clip]{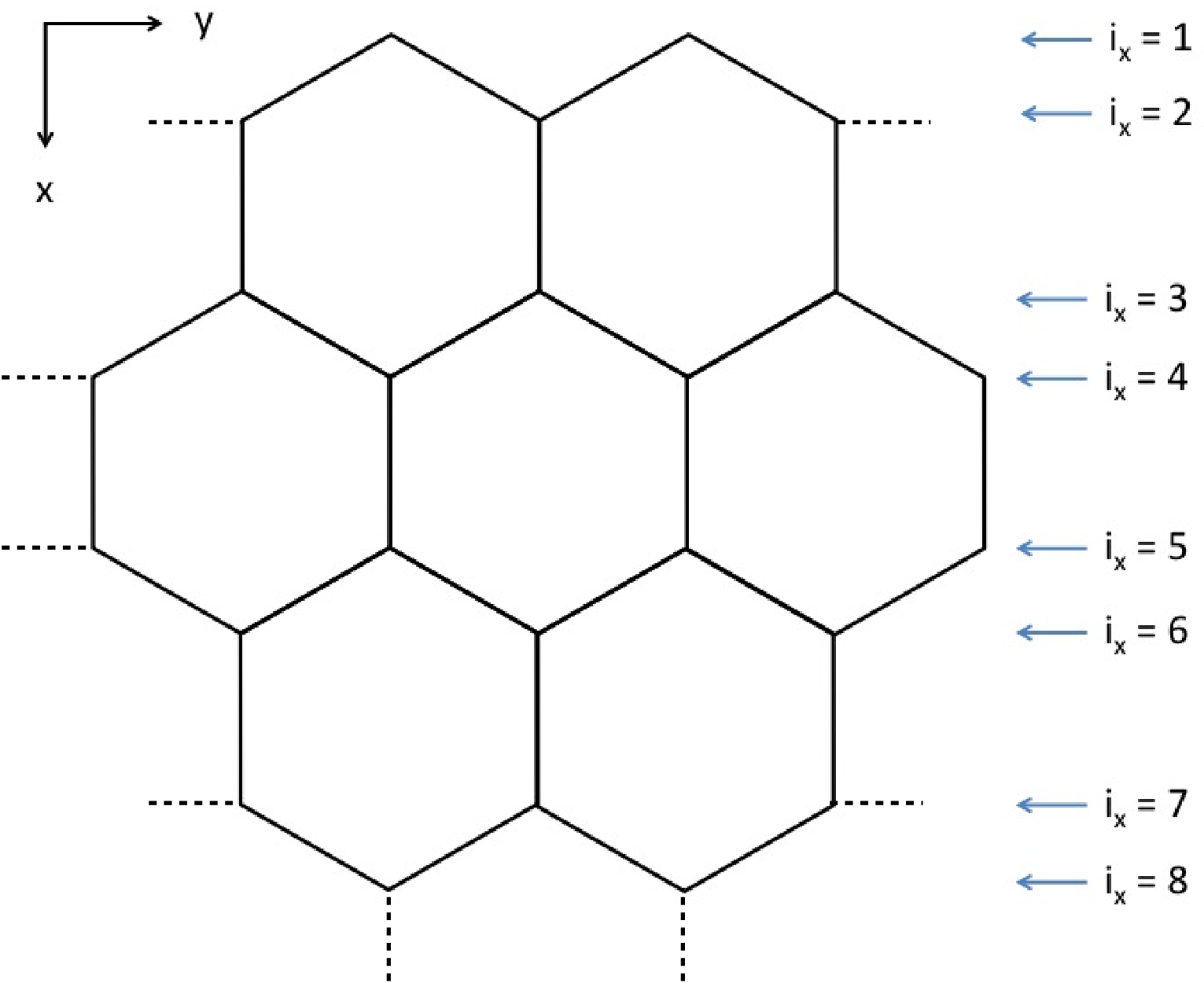}
\end{center}
\caption{
(Color online)
Schematic of a semi-infinite honeycomb lattice.
Dimers are located on the lattice points.
They are aligned along the $z$-direction.
$i_{x}=1,~2,~3$, etc., represent the $i_{x}$th sites.
A zigzag edge is at $i_x=1$.
There is translational symmetry along the $y$-direction.
}
\label{fig:honeycomb-edge}
\end{figure}

Let us consider a semi-infinite bilayer honeycomb lattice with a zigzag edge
at $i_x=1$ as shown in Fig. \ref{fig:honeycomb-edge}.
We represent the spin site by $i_{x}$ and $i_{y}$ along the $x$- and $y$-directions, respectively.
For numerical calculations, we introduce another zigzag edge at $i_x=N_x$.
The semi-infinite nature is reproduced with a sufficiently large $N_x$.
The model Hamiltonian is given by
\begin{align}
\H = \sum_{i} \H_{\rm intra}(i) + \sum_{\langle ij\rangle} \H_{\rm inter}(i,j),
\label{eqn:H-total-0}
\end{align}
with
\begin{align}
&\H_{\rm intra}(i) = J \bS_{il}\cdot\bS_{ir} - \bh\cdot ( \bS_{il} + \bS_{ir} ), \cr
&\H_{\rm inter}(i,j) = J' ( \bS_{il}\cdot\bS_{jl} + \bS_{ir}\cdot\bS_{jr} ).
\end{align}
We write $i\rightarrow i_{x}i_{y}$ and $j\rightarrow j_{x}j_{y}$ for simplicity.

\subsection{Mean-field theory}

The mean-field Hamiltonian for the $i$th site is expressed as
\begin{align}
\H_{\rm MF}(i) &= J \bS_{il}\cdot\bS_{ir} - \bh\cdot ( \bS_{il} + \bS_{ir} ) \cr
&~~~
+ J' \sum_{\gamma=l,r} \sum_{\langle j\rangle} \bM_{j\gamma}\cdot\bS_{i\gamma}.
\end{align}
Here, $\bM_{j\gamma}$ represents the magnetic moment on the $\gamma(=l,r)$ side of a dimer at the $j$th site.
The summation $\sum_{\langle j\rangle}$ is taken over the nearest-neighbors for the $i$th site.
The Hamiltonian is expressed in a $4\times 4$ matrix form.
Energy eigenstates are obtained by diagonalizing the mean-field Hamiltonian.
The magnetic moment is determined by
\begin{align}
\bM_{i\gamma} = \langle i0| \bS_{i\gamma} |i0\rangle,
\end{align}
where $|i0\rangle$ is the mean-field ground state at the $i$th site.
We solve the mean-field problem iteratively until a self-consistent solution for $\bM_{i\gamma}$
is obtained at each $i$th site.
Since the interdimer interaction is AF on the bipartite honeycomb lattice,
an AF ground state with a staggered moment is expected in the bulk region.
The ground state is affected by the solution in the bulk region
and we assume that the mean-field solution is independent of the position $i_{y}$.
In general, however, there is a possibility of a ground state
with broken translational invariance along the edge direction.
We did not explore this possibility in the present study.

\subsection{Magnetic excitation}

The magnetic excitation is obtained for the mean-field ground and excited states.
We write the energy eigenstate of the main-field Hamiltonian at the $i$th site as
\begin{align}
|im\rangle~~~(m=0,1,2,3).
\end{align}
Here, $m=0$ represents the mean-field ground state, while $m=1,2,3$ are those for the excited states.
The intradimer part of the Hamiltonian at the $i$th site is then expressed as
\begin{align}
&\H_{\rm intra}(i) = \sum_{m,n=0}^{3}
|im\rangle \langle im| \H_{\rm intra}(i) |in\rangle \langle in|.
\end{align}
At each site, we introduce the following boson, which creates the energy eigenstate out of the vacuum state:
\cite{Sachdev-1990}
\begin{align}
a_{im}^\dagger |{\rm vac}\rangle = |im\rangle.
\end{align}
The bosons are subjected to the following local constraint:
\begin{align}
\sum_{m=0}^3 a_{im}^\dagger a_{im} = 1.
\label{eqn:constraint-a}
\end{align}
$\H_{\rm intra}(i)$ is then expressed in terms of the bosons as
\cite{Shiina-2003}
\begin{align}
\H_{\rm intra}(i) = \sum_{m,n=0}^3 \H_{mn}^0(i) a_{im}^\dagger a_{in},
\label{eqn:H-i}
\end{align}
where $\H_{mn}^0(i)$ is the matrix element defined by
\begin{align}
\H_{mn}^0(i) = \langle im| \H_{\rm intra}(i) | in \rangle.
\end{align}
In the same manner, the spin operator is expressed as
\cite{Shiina-2003}
\begin{align}
\bS_{i\gamma} = \sum_{m,n=0}^3 \bS_{mn}(i,\gamma) a_{im}^\dagger a_{in},
\label{eqn:S-i}
\end{align}
with
\begin{align}
\bS_{mn}(i,\gamma) = \langle im| \bS_{i\gamma} | in \rangle.
\end{align}

Next, we replace the $a_{i0}$ operator for the mean-field ground state
with the local constraint [Eq. (\ref{eqn:constraint-a})] as
\cite{Sommer-2001,Matsumoto-2002,Matsumoto-2004}
\begin{align}
&a_{i0}^\dagger a_{i0} \rightarrow 1 - \sum_{m=1}^3 a_{im}^\dagger a_{im}, \cr
&a_{i0}^\dagger \rightarrow \left( 1 - \sum_{m=1}^3 a_{im}^\dagger a_{im} \right)^{1/2}, \cr
&a_{i0} \rightarrow \left( 1 - \sum_{m=1}^3 a_{im}^\dagger a_{im} \right)^{1/2}.
\label{eqn:holstein}
\end{align}
We substitute Eq. (\ref{eqn:holstein}) into Eqs. (\ref{eqn:H-i}) and (\ref{eqn:S-i})
and eliminate the $a_{i0}$ and $a_{i0}^\dagger$ operators.
$\H_{\rm intra}(i)$ and $\bS_{i\gamma}$ are expressed in terms of the bosons for the excited states.
The total Hamiltonian [Eq. (\ref{eqn:H-total-0})] is then written in the following form:
\begin{align}
\H = \H^{(0)} + \H^{(1)} + \H^{(2)} + \cdots.
\end{align}
Here, $\H^{(0)}$ is the c-number term that corresponds to the mean-field ground state energy.
The $\H^{(1)}$ term contains only one Bose operator of the excited state.
It vanishes when we use the appropriate self-consistently determined mean-field solution.
The $\H^{(2)}$ term contains two bosons.
We neglect higher-order terms of bosons since bosons are expected to be dilute at low temperatures.
The Hamiltonian is then expressed in the following form:
\begin{align}
\H = \H_{\rm local} + \H_{\rm nonlocal}.
\label{eqn:H-total-1}
\end{align}
Here, $\H_{\rm local}$ and $\H_{\rm nonlocal}$ are the Hamiltonian for the local and nonlocal parts defined by
\begin{align}
&\H_{\rm local} = \sum_{m,n=1}^3 \sum_i \Lambda_{mn}^i a_{im}^\dagger a_{in},
\label{eqn:H-total-2} \\
&\H_{\rm nonlocal} = \sum_{m,n=1}^3 \sum_{\langle ij \rangle}
\left(
  \Pi_{mn}^{ij} a_{im}^\dagger a_{jn} + \Delta_{mn}^{ij} a_{im}^\dagger a_{jn}^\dagger + {\rm h. c.}
\right), \nonumber
\end{align}
with
\begin{align}
&\Lambda_{mn}^i = \H_{mn}^{0}(i) - \H_{00}^{0}(i)\delta_{mn} \cr
&~~~~~~
+ J' \sum_\gamma \left[ \bS_{mn}(i,\gamma) - \bM_{i\gamma} \delta_{mn} \right] \cdot
      \sum_{\langle j\rangle} \bM_{j\gamma}, \cr
&\Pi_{mn}^{ij} = J' \sum_{\gamma} \bS_{m0}(i,\gamma) \cdot \bS_{0n}(j,\gamma), \cr
&\Delta_{mn}^{ij} = J' \sum_{\gamma} \bS_{m0}(i,\gamma) \cdot \bS_{n0}(j,\gamma).
\label{eqn:Lambda-Pi}
\end{align}
Here, the square root in Eq. (\ref{eqn:holstein}) is expanded when it is substituted into Eq. (\ref{eqn:S-i}).
In Eq. (\ref{eqn:Lambda-Pi}), we used the relation $\bS_{00}(i,\gamma)=\bM_{i\gamma}$.
In Eq. (\ref{eqn:H-total-2}), we omitted a constant term.
Note that $\Pi_{mn}^{ij}$ and $\Delta_{mn}^{ij}$ have the following relations:
\begin{align}
\Pi_{mn}^{ji} = \left( \Pi_{nm}^{ij} \right)^*,~~~~~~
\Delta_{mn}^{ji} = \Delta_{nm}^{ij}.
\label{eqn:relation}
\end{align}

Since there is a translational symmetry along the $y$-direction, as shown in Fig. \ref{fig:honeycomb-edge},
$\Lambda_{mn}^i$, $\Pi_{mn}^{ij}$, and $\Delta_{mn}^{ij}$ in Eq. (\ref{eqn:Lambda-Pi})
depend only on the positions of $i_x$ and $j_x$, i.e., they are independent of $i_y$ and $j_y$.
We then introduce the following Fourier transformation along the $y$-direction for the boson:
\begin{align}
a_{im} = \frac{1}{\sqrt{N_y}} \sum_k a_{ki_x m} e^{i k y_i}.
\label{eqn:Fourier-boson}
\end{align}
Here, $k$, $N_y$, and $y_i$ are the wavenumber, the number of dimer sites,
and the position of the dimer along the $y$-direction, respectively.
Substituting Eq. (\ref{eqn:Fourier-boson}) into Eq. (\ref{eqn:H-total-1}), we obtain
\begin{align}
\H &= \sum_k \sum_{m,n=1}^3 \sum_{i_x} \Lambda_{mn}^{i_x} a_{k i_x m}^\dagger a_{k i_x n} \cr
&~~~+ \sum_k \sum_{m,n=1}^3 \sum_{\langle i_x j_x \rangle} \left[
  \gamma_k^{i_xj_x} \left( \Pi_{mn}^{i_x j_x} a_{k i_x m}^\dagger a_{k j_x n} \right. \right. \cr
&~~~~~~~~~+
  \left. \left. \Delta_{mn}^{i_x j_x} a_{k i_x m}^\dagger a_{-k j_x n}^\dagger \right) + {\rm h. c.} \right].
\end{align}
Here, $\gamma_k^{i_xj_x}$ represents the momentum dependence for the excited states.
The specific forms of $\Lambda_{mn}^{i_x}$ and $\gamma_k^{i_x j_x}$ are given in Appendix \ref{sec:appendix-matrix}.

We next introduce the following $3N_x$-dimensional transposed vectors:
\begin{align}
&\ba_k^{\rm T} = \left( \cdots a_{k1m} \cdots a_{k2m} \cdots a_{kN_x m} \cdots \right), \cr
&{\ba_k^\dagger}^{\rm T}
= \left( \cdots a_{k1m}^\dagger \cdots a_{k2m}^\dagger \cdots a_{kN_x m}^\dagger \cdots \right).
\end{align}
Here, $N_x$ is the number of dimer sites along the $x$-direction.
Using these vectors, we introduce the following $6N_x$-dimensional vector:
\begin{align}
\vec{a}_k =
\left(
  \begin{array}{c}
    \ba_k \cr
    \ba_{-k}^\dagger
  \end{array}
\right).
\end{align}
Then, we obtain the following commutation relation:
\begin{align}
\H \vec{a}_k - \vec{a}_k \H = - \hat{\epsilon}_k \vec{a}_k.
\label{eqn:commutation}
\end{align}
Here, $\hat{\epsilon}_k$ is expressed as
\begin{align}
\hat{\epsilon}_k =
\left(
  \begin{array}{cc}
    \bA_k & \bB_k \cr
    - \bB_{-k}^* & - \bA_{-k}^*
  \end{array}
\right),
\label{eqn:epsilon}
\end{align}
with
\begin{align}
&\bA_k =
\left(
  \begin{array}{ccccc}
    \bLambda^{1} & \bPi_k^{12} & 0 & \cdots & 0 \cr
    \bPi_k^{21} & \bLambda^{2} & \bPi_k^{\rm 23} & \cdots & 0 \cr
    \vdots & \vdots & \vdots & \vdots & \vdots \cr
    0 & 0 & \cdots & \bLambda^{N_x-1} & \bPi_k^{N_x-1,N_x} \cr
    0 & 0 & \cdots & \bPi_k^{N_x,N_x-1} & \bLambda^{N_x}
  \end{array}
\right), \cr
&\bB_k =
\left(
  \begin{array}{ccccc}
    0 & \bDelta_k^{12} & 0 & \cdots & 0 \cr
    \bDelta_k^{21} & 0 & \bDelta_k^{\rm 23} & \cdots & 0 \cr
    \vdots & \vdots & \vdots & \vdots & \vdots \cr
    0 & 0 & \cdots & 0 & \bDelta_k^{N_x-1,N_x} \cr
    0 & 0 & \cdots & \bDelta_k^{N_x,N_x-1} & 0
  \end{array}
\right).
\label{eqn:matrix-AB}
\end{align}
Here, $\bLambda^{i_x}$, $\bPi_k^{i_x j_x}$, and $\bDelta_k^{i_x j_x}$ are $3\times 3$ matrices.
Their matrix elements are given by
\begin{align}
&\left(\bLambda^{i_x}\right)_{mn} = \Lambda_{mn}^{i_x}, \cr
&\left(\bPi_k^{i_x j_x}\right)_{mn} = \Pi_{mn}^{i_x j_x} \gamma_k^{i_x j_x}, \cr
&\left(\bDelta_k^{i_x j_x}\right)_{mn} = \Delta_{mn}^{i_x j_x} \gamma_k^{i_x j_x}.
\end{align}
In Eq. (\ref{eqn:matrix-AB}), we can see that the nearest-neighbor $i_x$ and $j_x$ sites are coupled,
reflecting the honeycomb lattice shown in Fig. \ref{fig:honeycomb-edge}.

Next, we assume that the Hamiltonian has the following diagonal form:
\begin{align}
\H = \sum_k E_k \alpha_k^\dagger \alpha_k.
\end{align}
Here, the $\alpha_k$ boson satisfies the commutation relation
\begin{align}
\H \alpha_k - \alpha_k \H = - E_k \alpha_k.
\end{align}
We then introduce the following Bogoliubov transformation:
\begin{align}
\alpha_k = \vec{X}_k^{\rm T} \vec{a}_k.
\label{eqn:Bogoliubov-0}
\end{align}
Here, $\vec{X}_k^{\rm T}$ is a $6N_x$-dimensional transposed vector.
Substituting Eq. (\ref{eqn:Bogoliubov-0}) into Eq. (\ref{eqn:commutation}), we reach the following eigenvalue equation:
\begin{align}
\hat{\epsilon}_k^{\rm T} \vec{X}_k = E_k \vec{X}_k.
\label{eqn:eigen}
\end{align}
When we diagonalize the $6N_x\times 6N_x$ matrix $\hat{\epsilon}_k^{\rm T}$,
the energy eigenvalues are obtained as pairs of $\pm E_k$.
Here, $+$ and $-$ correspond to the particle and hole solutions, respectively.
We only take the particle (positive eigenvalue) solution.
There are $3N_x$ excitation modes in the Brillouin zone reflecting the $N_x$ sites since each dimer provides three states.

\section{Disordered Phase}
\label{sec:analytic-dimer}

In the disordered phase, the singlet and triplet states are the mean-field solutions.
This enables us to treat the problem analytically.
At each dimer site, the spin operators are expressed as Eq. (\ref{eqn:spin-operator}).
Since we retain the triplet excitation up to the second order,
the singlet operator can be rewritten with $s_{i}\rightarrow 1$ and $s_{i}^\dagger\rightarrow 1$
in the interdimer interaction part of the Hamiltonian.
The Hamiltonian is then expressed as Eq. (\ref{eqn:Hij}).
The Hamiltonian is divided into two parts as $\H=\H_0+\H_{\pm}$.
The former and latter contain the $t_{i0}$ and $t_{i\pm}$ bosons, respectively.
The bosons appear in the $S^z=0$ and $S^z=\pm 1$ excited triplet states, respectively.
For both $\H_0$ and $\H_\pm$ in Eq. (\ref{eqn:Hij}),
the first term represents the local energy of the triplet state under a finite field $h$.
The second term represents the hopping of the triplet excitation.
The last term represents the pair creation and annihilation of the triplet excitation.
In the case of graphene, the last term does not appear in the Hamiltonian.
Thus, the existence of the both terms in the Hamiltonian is peculiar to the spin dimer system.
Since the Hamiltonian is divided into two parts,
we discuss the eigenvalue problem of Eq. (\ref{eqn:eigen}) separately in terms of $\H_{0}$ and $\H_\pm$.

\subsection{$\H_{0}$ part}

In this study, we assume a semi-infinite system along the $x$-direction,
i.e., there is only one zigzag edge at $i_{x}=1$ in Fig. \ref{fig:honeycomb-edge}.
For $\H_{0}$, the matrices $\bA_{k}$ and $\bB_{k}$ in Eq. (\ref{eqn:epsilon}) are written as
\begin{align}
\bA_{k} = J + \bm{M}_{k},~~~~~~
\bB_{k} = \bm{M}_{k},
\label{eqn:AB}
\end{align}
where $\bm{M}_{k}$ is given by
\begin{align}
\bm{M}_{k} =
\left(
\begin{matrix}
0 & f & 0 & 0 & 0 & 0 & \cdots \\
f & 0 & g & 0 & 0 & 0 & \cdots \\
0 & g & 0 & f & 0 & 0 & \cdots \\
0 & 0 & f & 0 & g & 0 & \cdots \\
\vdots & \vdots & \vdots & \vdots & \vdots & \vdots & \vdots
\label{eqn:M}
\end{matrix}
\right),
\end{align}
with
\begin{align}
f = J' \cos\left(\frac{k}{2}\right),~~~
g = \frac{1}{2} J'.
\end{align}
The eigenvalue equation [Eq. (\ref{eqn:eigen})] is then written as
\begin{align}
\left(
\begin{matrix}
J+\bm{M}_{k} & -\bm{M}_{k} \\
\bm{M}_{k} & -J-\bm{M}_{k}
\end{matrix}
\right)
\left(
\begin{matrix}
\bm{\psi}_k \\
\bm{\bar{\psi}}_k
\end{matrix}
\right)
= E
\left(
\begin{matrix}
\bm{\psi}_k \\
\bm{\bar{\psi}}_k
\end{matrix}
\right),
\label{eqn:eigen-flat}
\end{align}
where we write
\begin{align}
\vec{X}_{k} =
\left(
\begin{matrix}
\bm{\psi}_{k} \cr
\bm{\bar{\psi}}_{k}
\end{matrix}
\right)
\label{eqn:X1}
\end{align}
and used the following relations:
\begin{align}
\bm{A}_{k} = \bm{A}_{-k} = \bm{A}_{k}^{\rm T},~~~~~~
\bm{B}_{k} = \bm{B}_{-k} = \bm{B}_{k}^{\rm T}.
\end{align}
In Eq. (\ref{eqn:eigen-flat}), $\bm{\psi}_{k}$ and $\bm{\bar{\psi}}_{k}$ are the field operators
for the particle and hole, respectively.

Note that the matrix $\bm{M}_{k}$ has the same form as that for graphene.
\cite{Fujita-1996}
As in Eq. (\ref{eqn:AB}), $\bm{M}_k$ appears in both $\bm{A}_k$ and $\bm{B}_k$.
Here, $\bm{M}_k$ in $\bm{A}_k$ and $\bm{B}_k$ represents the hopping
and pair creation and annihilation processes for the triplet state, respectively.
This originates from the fact that the excited triplet state can move via both processes.
In the case of graphene, however, there is no $\bm{B}_k$ matrix.
\cite{Fujita-1996}
On the other hand, $\bm{M}_k$ only appears in $\bm{B}_k$ in case of an AF spin system on a monolayer honeycomb lattice
since the excitation can move only by the pair creation and annihilation process.
\cite{You-2008}
Note that the spin dimer system has both terms.
In this sense, the spin dimer model connects the graphene and the AF spin system on a monolayer honeycomb lattice.

Following Ref. \ref{ref:You}, we analyze the bulk and edge magnetic excitations.
The eigenvectors for the particle and hole can be written as
\begin{align}
\bm{\psi}_{k} =
\left(
\begin{matrix}
a_{1} \cr
b_{1} \cr
a_{2} \cr
b_{2} \cr
\vdots
\end{matrix}
\right),~~~~~~
\bm{\bar{\psi}}_{k} =
\left(
\begin{matrix}
\bar{a}_{1} \cr
\bar{b}_{1} \cr
\bar{a}_{2} \cr
\bar{b}_{2} \cr
\vdots
\end{matrix}
\right).
\label{eqn:X2}
\end{align}
Substituting Eq. (\ref{eqn:X2}) into Eq. (\ref{eqn:eigen-flat}), we obtain
\begin{align}
Ja_{n} + g b_{n-1} + f b_{n} - g \bar{b}_{n-1} - f \bar{b}_{n} &= E a_{n}, \cr
Jb_{n} + f a_{n} + g a_{n+1} - f \bar{a}_{n} - g \bar{a}_{n+1} &= E b_{n}, \cr
-J\bar{a}_{n} - g \bar{b}_{n-1} - f \bar{b}_{n} + g b_{n-1} + f b_{n} &= E \bar{a}_{n}, \cr
-J\bar{b}_{n} - f \bar{a}_{n} - g \bar{a}_{n+1} + f a_{n} + g a_{n+1} &= E \bar{b}_{n}.
\label{eqn:abcd}
\end{align}
At the edge site, there is no nearest-neighbor site along the negative $x$-direction.
Thus, the equations at the edge site should be
\begin{align}
Ja_{1} + f b_{1} - f \bar{b}_{1} &= E a_{1}, \cr
-J\bar{a}_{1} - f \bar{b}_{1} + f b_{1} &= E \bar{a}_{1}.
\label{eqn:condition-boundary}
\end{align}

First, we study the bulk solution.
For the semi-infinite system, the energy region for the bulk excitation is not affected by the boundary condition.
Therefore, we introduce a periodic condition along the $x$-direction and solve the eigenvalue equation by assuming
\begin{align}
\left(
\begin{matrix}
a_{n} \cr
b_{n} \cr
\bar{a}_{n} \cr
\bar{b}_{n}
\end{matrix}
\right)=
\left(
\begin{matrix}
a_0 \cr
b_0 \cr
\bar{a}_0 \cr
\bar{b}_0
\end{matrix}
\right) z^{n}.
\label{eqn:zn}
\end{align}
For extended states, $z$ is a complex number such as
\begin{align}
z=e^{\pm i k_{x}}.
\label{eqn:z-k}
\end{align}
Here, $k_{x}$ plays the role of a wavevector along the $x$-direction.
Substituting Eq. (\ref{eqn:zn}) into Eq. (\ref{eqn:abcd}), we obtain
\begin{align}
\left(
\begin{matrix}
J-E & f+\frac{g}{z} & 0 & -f-\frac{g}{z} \cr
f+zg & J-E & -f-zg & 0 \cr
0 & f+\frac{g}{z} & -J-E & -f-\frac{g}{z} \cr
f+zg & 0 & -f-zg & -J-E
\end{matrix}
\right)
\left(
\begin{matrix}
a_0 \cr
b_0 \cr
\bar{a}_0 \cr
\bar{b}_0
\end{matrix}
\right) = 0.
\label{eqn:eigen-bulk}
\end{align}
The energy eigenvalues are given by
\begin{align}
E = \pm \left[ J \pm 2 \sqrt{ f^{2} + g^{2} + fg \left( z + \frac{1}{z} \right) } \right].
\label{eqn:E-bulk}
\end{align}
Here, the positive and negative energies are for the particle and hole solutions, respectively.
Equation (\ref{eqn:z-k}) leads to $z+\frac{1}{z}=2\cos(k_{x})$, which varies in the range
\begin{align}
-2 \le z+\frac{1}{z} \le 2.
\label{eqn:z-condition}
\end{align}
Substituting Eq. (\ref{eqn:z-condition}) into Eq. (\ref{eqn:E-bulk}),
we obtain the following energies, which determine the boundary surrounding the bulk excitations:
\begin{align}
&E_{k,1\pm} = \sqrt{ J^{2} + JJ' \left| 2\cos\left(\frac{k}{2}\right) \pm 1 \right| }, \cr
&E_{k,2\pm} = \sqrt{ J^{2} - JJ' \left| 2\cos\left(\frac{k}{2}\right) \pm 1 \right| }.
\label{eqn:boundary}
\end{align}

We next study the edge solution taking the boundary condition into account.
Comparing Eqs. (\ref{eqn:condition-boundary}) and (\ref{eqn:abcd}) for $n=1$,
we notice that $b_{0}=\bar{b}_{0}$ must be satisfied in Eq. (\ref{eqn:abcd}) for $n=1$.
Substituting this relation in Eq. (\ref{eqn:eigen-bulk}),
we obtain the following energy eigenvalue and eigenvector:
\begin{align}
E=J,~~~a_0\neq 0,~~~b_0=\bar{a}_0=\bar{b}_0=0,
\label{eqn:E-flat}
\end{align}
with
\begin{align}
z=-\frac{1}{2\cos\left(\frac{k}{2}\right)}.
\label{eqn:z-zigzag}
\end{align}
This result indicates that an edge magnon mode with a completely flat dispersion relation emerges at $E=J$
in the presence of the zigzag edge.
In addition, $\bar{a}_{0}=\bar{b}_{0}=0$ means that the pair creation and annihilation term
does not contribute to the wave function of the edge state.
For a bound state, the wavefunction must decrease with increasing $n$ in Eq. (\ref{eqn:zn}).
To realize this, $z$ must satisfy $|z|<1$.
This means that the wavefunction of the edge mode decreases exponentially in the bulk region
and leads to the following condition:
\begin{align}
|k| > \frac{2\pi}{3}.
\label{eqn:k-condition}
\end{align}
Thus, the solution of the flat edge mode for the dimer system is exactly the same as that for the case of graphene.
\cite{Fujita-1996}

\subsection{$\H_{\pm}$ part}

\begin{figure}[t]
\begin{center}
\includegraphics[width=6cm,clip]{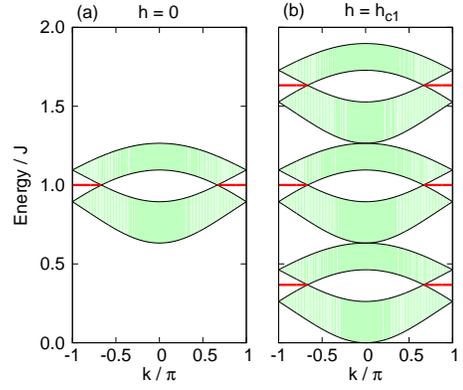}
\end{center}
\caption{
(Color online)
Energy region of the bulk magnon excitation and edge magnon mode for $J'/J=0.2$ under various fields.
(a) For $h=0$.
(b) For $h=h_{\rm c1}$.
The filled area represents the energy region for the bulk excitation.
The boundary lines are plotted with Eq. (\ref{eqn:boundary}).
The thick (red) line represents the flat edge magnon mode
that appears in the $|k|> \frac{2\pi}{3}$ region in the presence of the zigzag edge.
}
\label{fig:magnon-para}
\end{figure}

For $\H_{\pm}$, the matrices $\bA_{k}$ and $\bB_{k}$ in Eq. (\ref{eqn:epsilon}) are double in size
compared with the case of $\H_{0}$ since the Hamiltonian contains both the $S^{z}=\pm 1$ triplet states.
The matrices $\bA_{k}$ and $\bB_{k}$ are then written as
\cite{hopping-pm}
\begin{align}
&\bA_{k} =
\left(
\begin{array}{c|c}
J-h + \bm{M}_{k} & 0 \\
\hline
0 & J+h + \bm{M}_{k}
\end{array}
\right), \cr
&\bB_{k} =
\left(
\begin{array}{c|c}
0 & -\bm{M}_{k} \\
\hline
-\bm{M}_{k} & 0
\end{array}
\right).
\end{align}
Here, the basis functions for the matrix elements are separated by the $S^{z}=\pm 1$ triplet states.
Then, we write the eigenvector as
\begin{align}
\vec{X}_{k} =
\left(
\begin{matrix}
\bm{\psi}_{k+} \cr
\bm{\psi}_{k-} \cr
\bm{\bar{\psi}}_{k+} \cr
\bm{\bar{\psi}}_{k-}
\end{matrix}
\right).
\label{eqn:X-pm}
\end{align}
The eigenvalue equation is then separated into the following two parts:
\begin{align}
\left(
\begin{matrix}
J \mp h+\bm{M}_{k} & \bm{M}_{k} \\
-\bm{M}_{k} & -J \mp h-\bm{M}_{k}
\end{matrix}
\right)
\left(
\begin{matrix}
\bm{\psi}_{k\pm} \\
\bm{\bar{\psi}}_{k\mp}
\end{matrix}
\right)
= E
\left(
\begin{matrix}
\bm{\psi}_{k\pm} \\
\bm{\bar{\psi}}_{k\mp}
\end{matrix}
\right).
\label{eqn:eigen-flat-pm}
\end{align}
Since Eq. (\ref{eqn:eigen-flat-pm}) is essentially the same as Eq. (\ref{eqn:eigen-flat}),
note that the energy eigenvalue of Eq. (\ref{eqn:eigen-flat-pm}) is simply shifted by $\mp h$ from that for $\H_0$.

In Fig. \ref{fig:magnon-para}, we show the bulk and flat edge excitations for $h=0$ and $h=h_{\rm c1}$.
The bulk excitations are inside the filled area, whose boundary is plotted with Eq. (\ref{eqn:boundary}).
In the presence of the zigzag edge, an additional flat edge mode emerges at $E=J$.
At $h=0$, the triplet excitation is threefold degenerate.
Under a finite field, it splits into three modes.
At the lower critical field ($h=h_{\rm c1}$), the lowest excitation mode becomes soft.
We verified that the present result for the semi-infinite system is reproduced
within a finite-size numerical calculation for a large system size of $N_{x}$.

\section{Ordered Phase}

In the ordered phase, a finite magnetic moment appears
and generates a nonuniform mean-field potential for the magnetic excitation.
This may affect the edge magnon mode.
In this section, we study how the excited state changes in the magnetically ordered phase.

In the case of a spin dimer, the ordered moment shrinks even at zero temperature.
In addition, the moment size depends on the distance from the edge.
Since it is difficult to treat the magnetic excitation analytically in the ordered phase for the spin dimer system,
we perform a finite-size numerical calculation in this section.

\subsection{Magnetic field-induced order}

\begin{figure}[t]
\begin{center}
\includegraphics[width=7cm,clip]{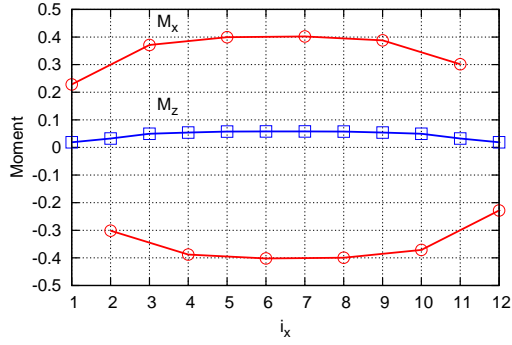}
\end{center}
\caption{
(Color online)
Position dependence of the self-consistently determined ordered moment on the left side of the dimers for $h=1.1 h_{\rm c1}$.
The squares and circles represent the ferromagnetic ($M_z$) and AF ($M_x$) components, respectively,
which are renormalized by the saturated moment.
Here, the $x$-axis is taken along the AF moment.
On the right side of the dimers, the sign of $M_x$ is reversed, while $M_z$ is the same.
The parameters are chosen as $J'=0.2J'_{\rm c}$ and $N_x=12$.
\label{fig:moment}
}
\end{figure}

\begin{figure}[t]
\begin{center}
\includegraphics[width=8cm,clip]{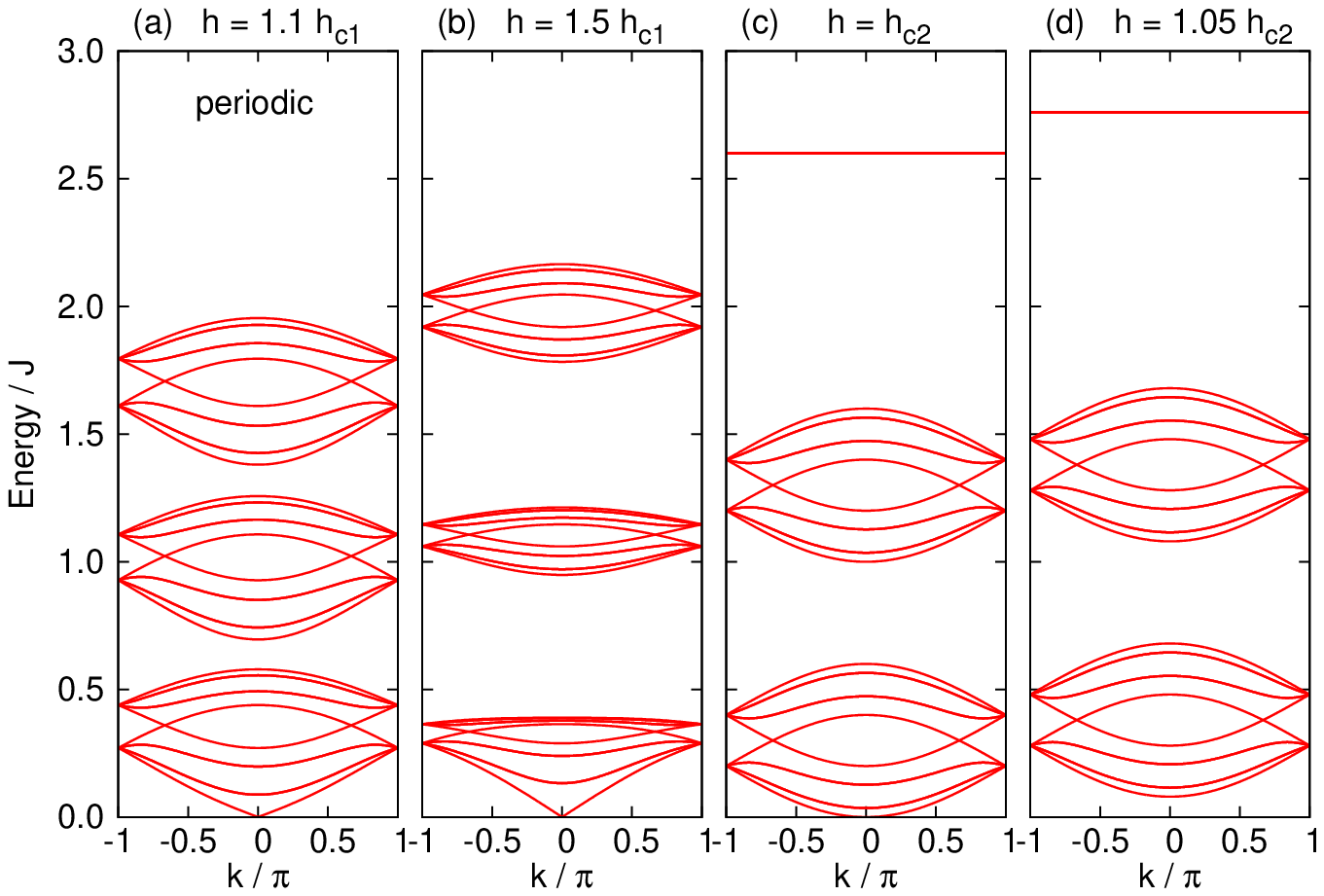}
\includegraphics[width=8cm,clip]{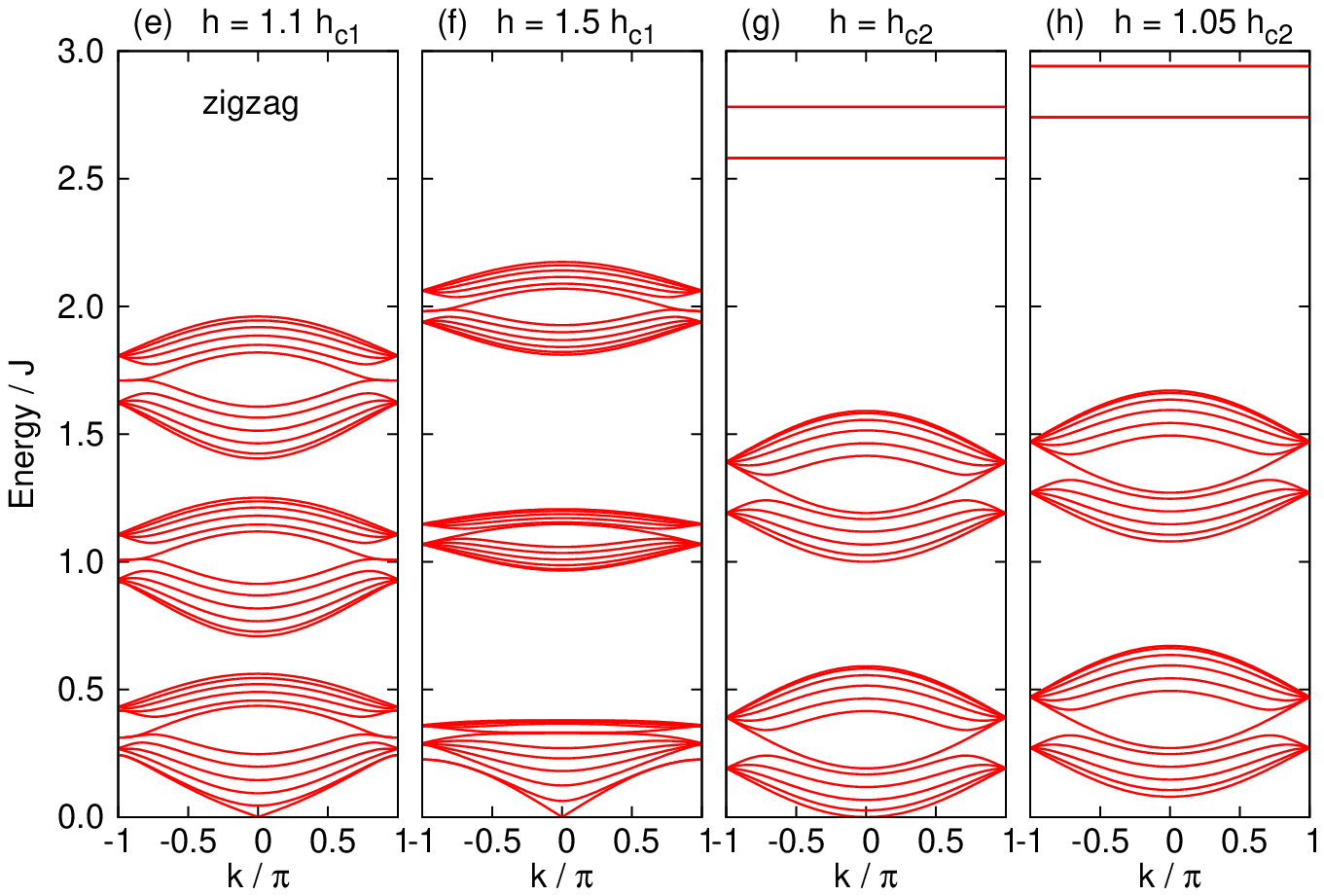}
\end{center}
\caption{
(Color online)
Magnon dispersion relation for various fields.
(a) For $h=1.1h_{\rm c1}$.
\cite{hc}
(b) For $h=1.5h_{\rm c1}$.
(c) For $h=h_{\rm c2}$.
(d) For $h=1.05h_{\rm c2}$.
(a)-(d) show the results for a periodic boundary condition along the $x$-direction.
(e) For $h=1.1h_{\rm c1}$.
(f) For $h=1.5h_{\rm c1}$.
(g) For $h=h_{\rm c2}$.
(h) For $h=1.05h_{\rm c2}$.
(e)-(h) show the results for a zigzag edge.
The parameters are chosen as $J'=0.2J$ and $N_x=12$.
\label{fig:energy-h}
}
\end{figure}

We first show the self-consistently determined ordered magnetic moment in Fig. \ref{fig:moment}
for a system size of $N_{x}=12$.
Here, we choose a moderate value of $N_{x}$ to show the excitation mode clearly.
Note that it is easy to treat a much larger system size.
In the field-induced ordered phase, there are both ferromagnetic and AF moments.
We can see that the moments at the edge sites ($i_x=1,12$) are reduced from those for the bulk sites.
This is owing to the fact that the number of nearest-neighbor sites is two for the edge sites,
while it is three for the bulk sites.
This suppresses the moment size by enhancing the singlet component of the mean-field ground state at the edge sites.

Next, we show the magnon dispersion relation under various fields in Fig. \ref{fig:energy-h}
for both periodic and zigzag-edge boundary conditions.
When the field is increased, the middle and higher branches are shifted to the high-energy region by the Zeeman effect.
In the ordered phase ($h_{\rm c1}<h<h_{\rm c2}$), the lowest mode shows a linear dispersion relation around $k=0$,
reflecting the Nambu-Goldstone mode in the AF ordered state.
In the saturated phase ($h > h_{\rm c2}$),
the lowest mode shows a quadratic dispersion relation with the opening of an excitation gap.
For the zigzag edge, we notice that the shape of the edge mode
deviates from the flat dispersion relation in the ordered phase [see Fig. \ref{fig:energy-h}(e)].
The deviation is distinct in the lower branch.
When the field is increased, the flat mode merges with the other modes
and the lowest mode separates from the others as shown in Fig. \ref{fig:energy-h}(f).
This originates from the appearance of a finite magnetic moment in the ordered phase.
The finite moment induces a mean-field potential via the interdimer interaction.
At the edge site ($i_x=1$), the mean-field potential is different from that for the other ($i_x\neq 1$) sites.
This is because there are only two nearest-neighbor sites at the edge.
This potential difference arises in the ordered phase and it affects the shape of the dispersion relation of the edge mode.
Note that this behavior is also seen in the case of graphene when we introduce a nonuniform potential.
\cite{Yao-2009}
Therefore, in the case of a dimer, the completely flat edge mode only appears in the disordered phase.
In the saturated phase for $h\ge h_{\rm c2}$, there are two kinds of excitations above 2.5 
[see Figs. \ref{fig:energy-h}(g) and \ref{fig:energy-h}(h)].
The lower is from a local excitation at the edge, while the higher is a local excitation from the bulk sites.

\begin{figure}[t]
\begin{center}
\includegraphics[width=8cm,clip]{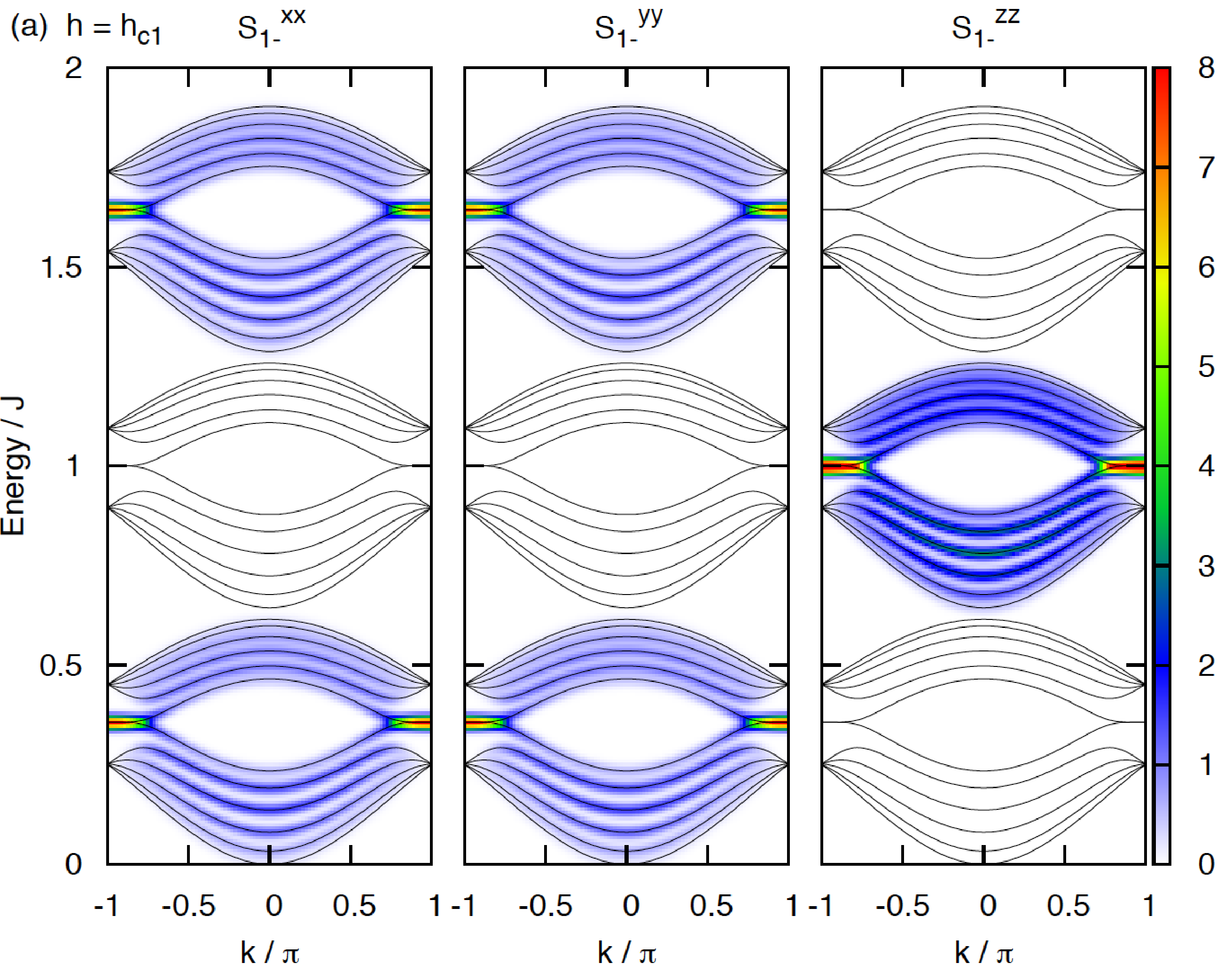}
\includegraphics[width=8cm,clip]{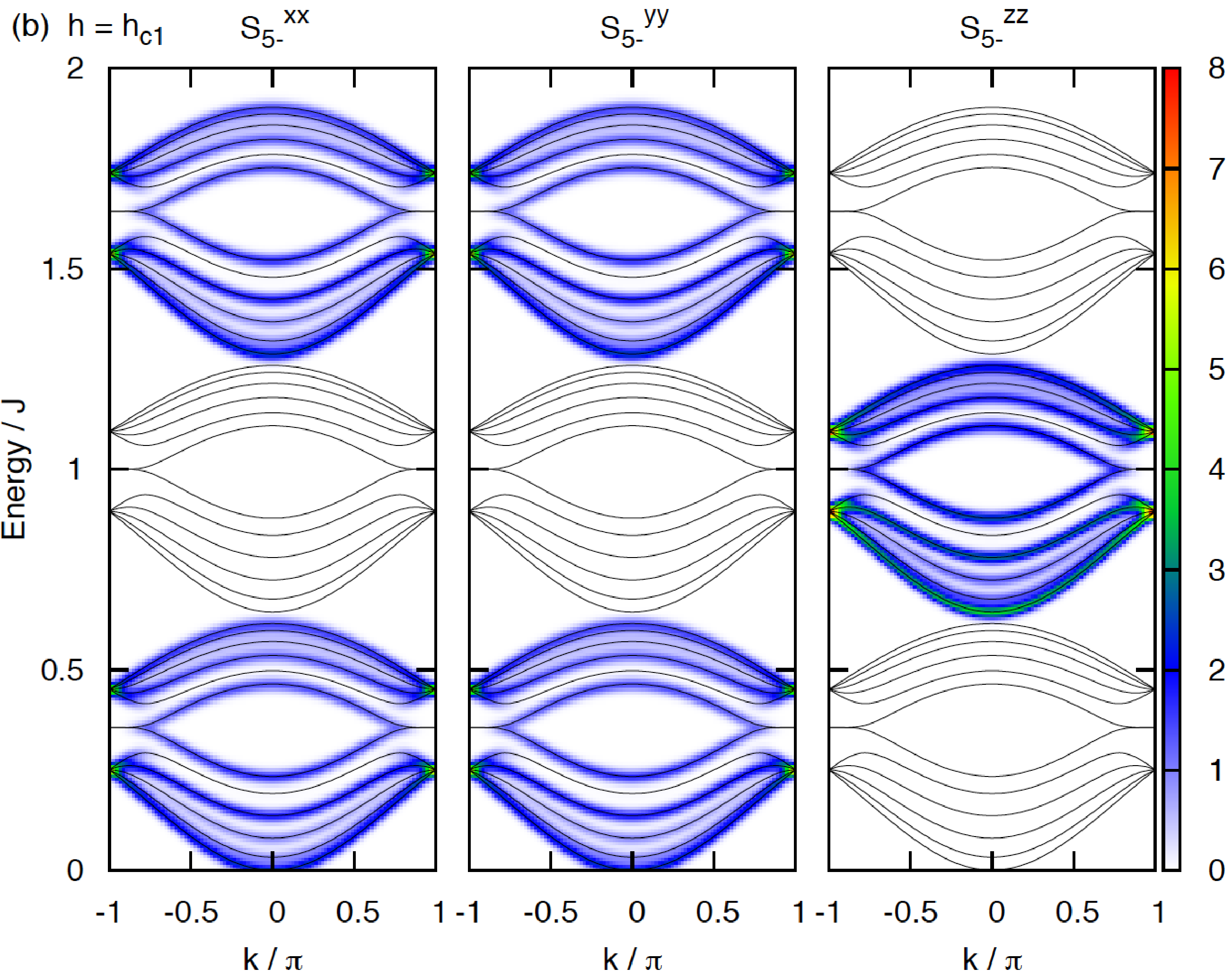}
\end{center}
\caption{
(Color online)
Dynamical spin correlation function defined by Eq. (\ref{eqn:dynamical-spin}) at the critical field $h_{\rm c1}$.
\cite{hc}
The thin black line represents the magnon dispersion relation.
The parameters are chosen as $J'=0.2J$ and $N_x=12$.
(a) For the edge site $(i_x=1)$.
(b) For the bulk site $(i_x=5)$.
\label{fig:sq-hc}
}
\end{figure}

\begin{figure}[t]
\begin{center}
\includegraphics[width=8cm,clip]{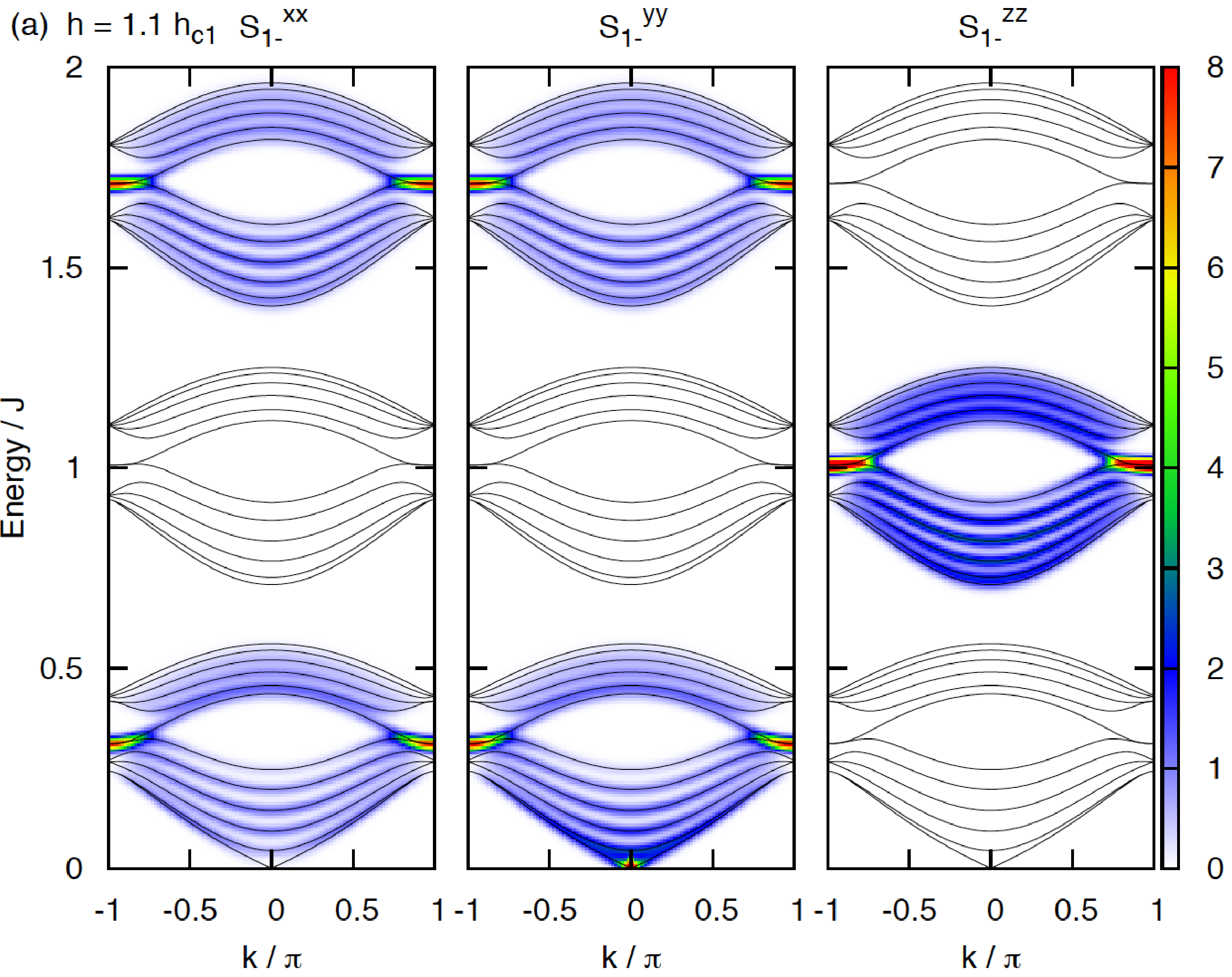}
\includegraphics[width=8cm,clip]{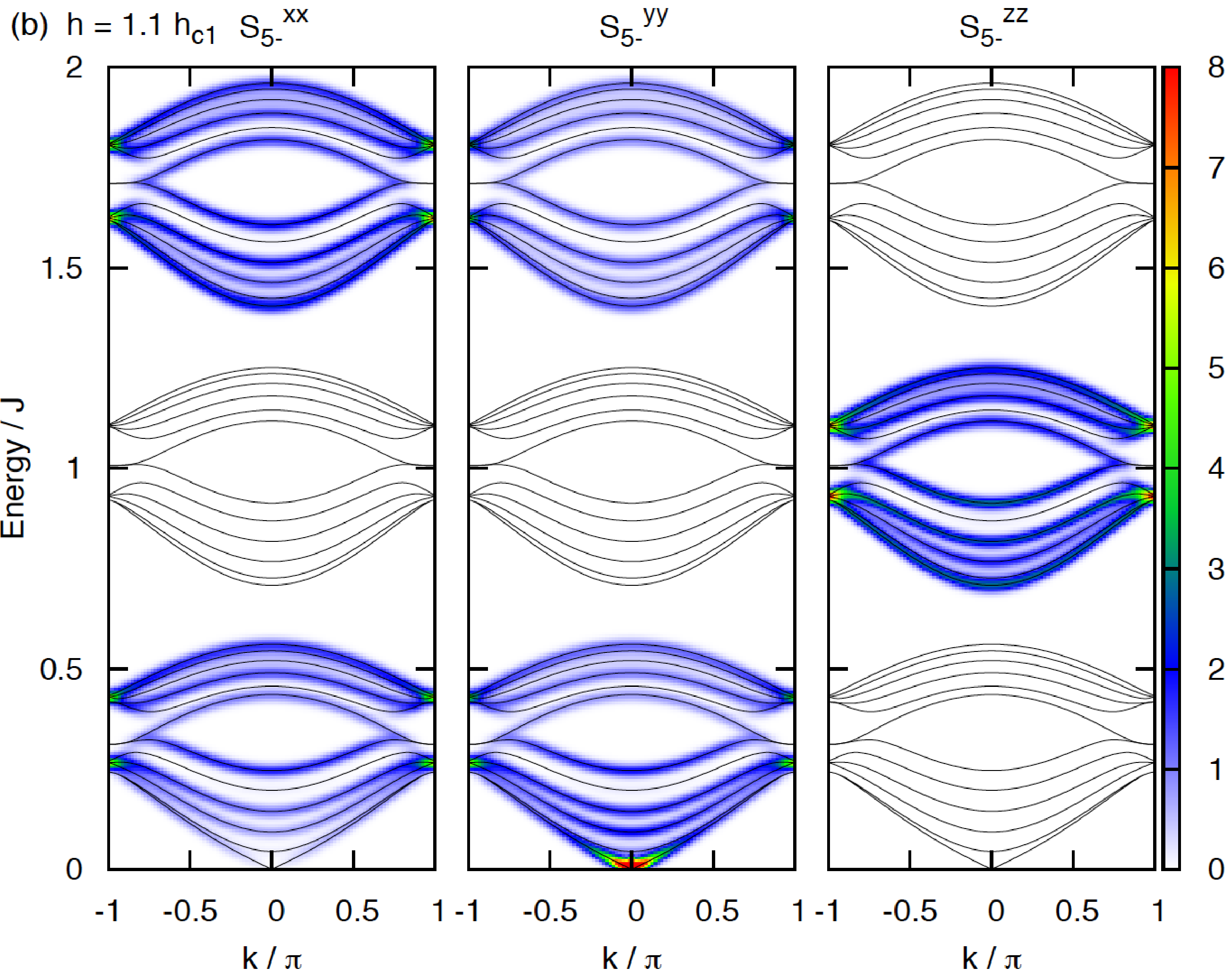}
\end{center}
\caption{
(Color online)
Dynamical spin correlation function in the ordered phase for $h=1.1h_{\rm c1}$.
A ferromagnetic moment appears along the field ($z$-direction), while an AF moment appears perpendicular to the field.
We take the $x$-direction along the AF moment.
(a) For the edge site $(i_x=1)$.
(b) For the bulk site $(i_x=5)$.
\label{fig:sq-1.1-hc}
}
\end{figure}

\begin{figure}[t]
\begin{center}
\includegraphics[width=8cm,clip]{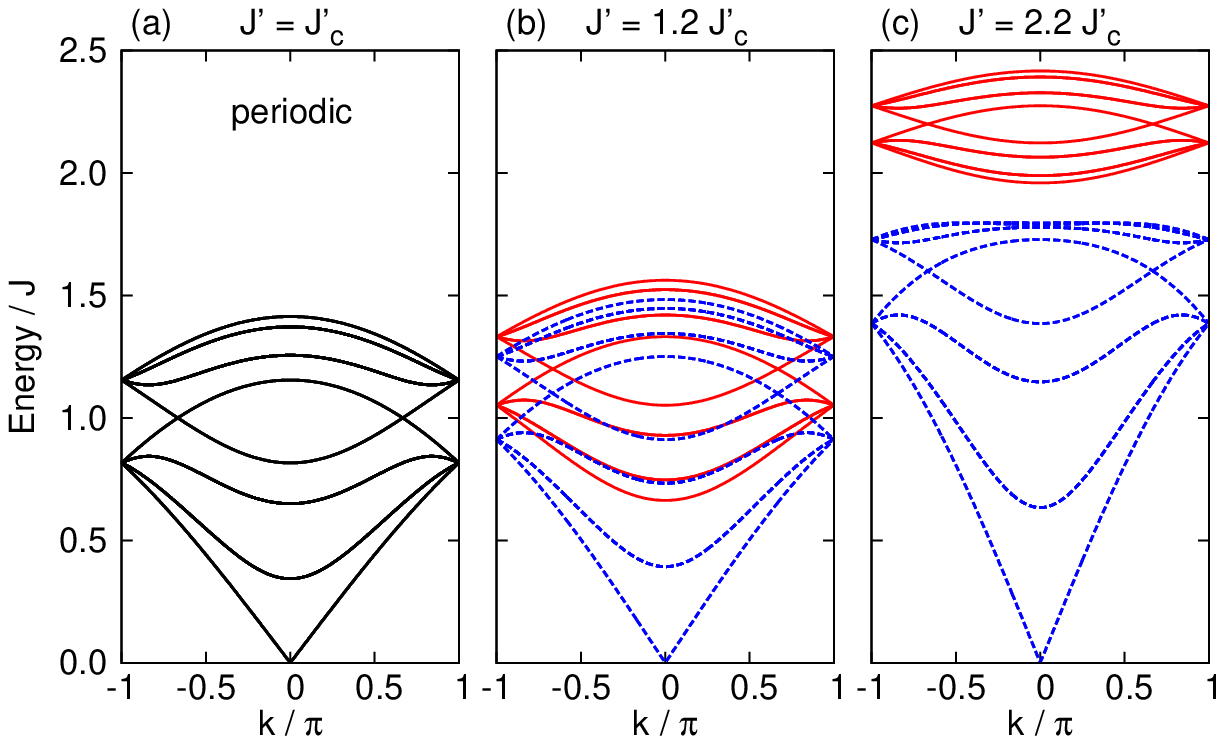}
\includegraphics[width=8cm,clip]{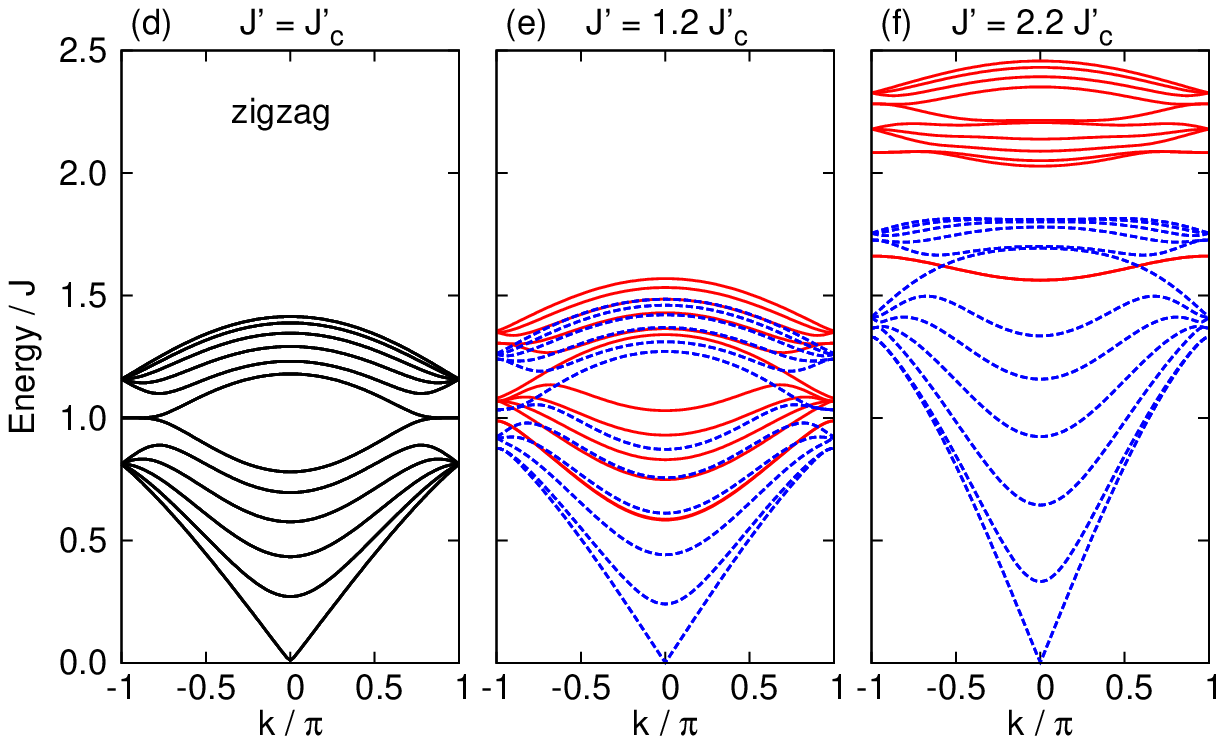}
\end{center}
\caption{
(Color online)
Magnon dispersion relation for various values of $J'$ under $h=0$.
(a) For $J'=J'_{\rm c}$.
(b) For $J'=1.2J'_{\rm c}$.
(c) For $J'=2.2J'_{\rm c}$.
(a)-(c) show the results for the periodic boundary condition along the $x$-direction.
(d) For $J'=J'_{\rm c}$.
\cite{jpc}
Each mode is threefold degenerate.
(e) For $J'=1.2J'_{\rm c}$.
(f) For $J'=2.2J'_{\rm c}$.
(d)-(f) show the results for the zigzag edge.
In the ordered phase, the solid (red) and dashed (blue) lines represent the L- and T-modes, respectively.
\label{fig:energy-jp-order}
}
\end{figure}

We next show the dynamical spin correlation function,
where the formulation is given in Appendix \ref{appendix-correlation}.
In the disordered phase, one-magnon excitation is possible between the singlet-triplet states.
In the ordered phase, the transition between the triplet-triplet states also participates in the process.
In the vicinity of the quantum critical point, however, the latter contribution is smaller
and we only consider the $S_{i_x-}^{\alpha\alpha}$ component (singlet-triplet process)
defined by Eq. (\ref{eqn:dynamical-spin}).
Figure \ref{fig:sq-hc} shows the dynamical spin correlation function at the critical field $h=h_{\rm c1}$.
Since the magnetic field is applied along the $z$-direction, the $x$- and $y$-directions are equivalent
($S_{i_x-}^{xx}=S_{i_x-}^{yy}$) for $h\le h_{\rm c1}$.
There are three kinds of excitation modes, reflecting the $S^z=1,0$, and $-1$ states under the field.
The lowest and highest branches have intensities in the correlation functions $S_{i_x-}^{xx}$ and $S_{i_x-}^{yy}$,
while the middle one has an intensity in the correlation function $S_{i_x-}^{zz}$.
Figure \ref{fig:sq-hc}(a) shows the contribution from the edge site ($i_x=1$).
We can see strong intensity on the flat dispersion.
As discussed in the previous section, the wavefunction of the flat mode is present only at the odd $i_x$ sites
[see $b_{0}=0$ in Eq. (\ref{eqn:E-flat})] and it decreases exponentially with increasing $i_x$.
Thus, we can see lower intensity on the flat mode for $i_x=5$ as shown in Fig. \ref{fig:sq-hc}(b).

We show the result for the ordered phase in Fig. \ref{fig:sq-1.1-hc}.
We can see that the edge mode becomes dispersive.
Although the mode loses the flat nature in its dispersion, we can see that it still keeps the bound state feature
by comparing Figs. \ref{fig:sq-1.1-hc}(a) and \ref{fig:sq-1.1-hc}(b).
In the ordered phase, the appearance of the ordered moment lifts the degeneracy in the $x$- and $y$-directions,
where the former and latter are along and perpendicular to the AF moment, respectively.
Since the correlation function $S_{i_x -}^{yy}$ corresponds to the transverse fluctuation of the AF moment,
it has a stronger intensity around $k=0$.
This point is clear on the bulk site [see Fig. \ref{fig:sq-1.1-hc}(b)].

\subsection{Interdimer interaction-induced order}

We first show the result for the periodic boundary condition
in Figs. \ref{fig:energy-jp-order}(a)-\ref{fig:energy-jp-order}(c).
In the ordered phase, the appearance of the AF moment lifts the threefold degeneracy
and the modes split into the L- and T-modes as shown in Fig. \ref{fig:energy-jp-order}(b).
The band width of the T-mode increases with $J'$, while it decreases for the L-mode.
For a large $J'$, the L-mode shifts to the high-energy region as shown in Fig. \ref{fig:energy-jp-order}(c).
In the presence of the zigzag edge, the flat edge mode appears [see Fig. \ref{fig:energy-jp-order}(d)].
In the ordered phase, the edge mode becomes dispersive as shown in Fig. \ref{fig:energy-jp-order}(e).

\begin{figure}[t]
\begin{center}
\includegraphics[width=7cm,clip]{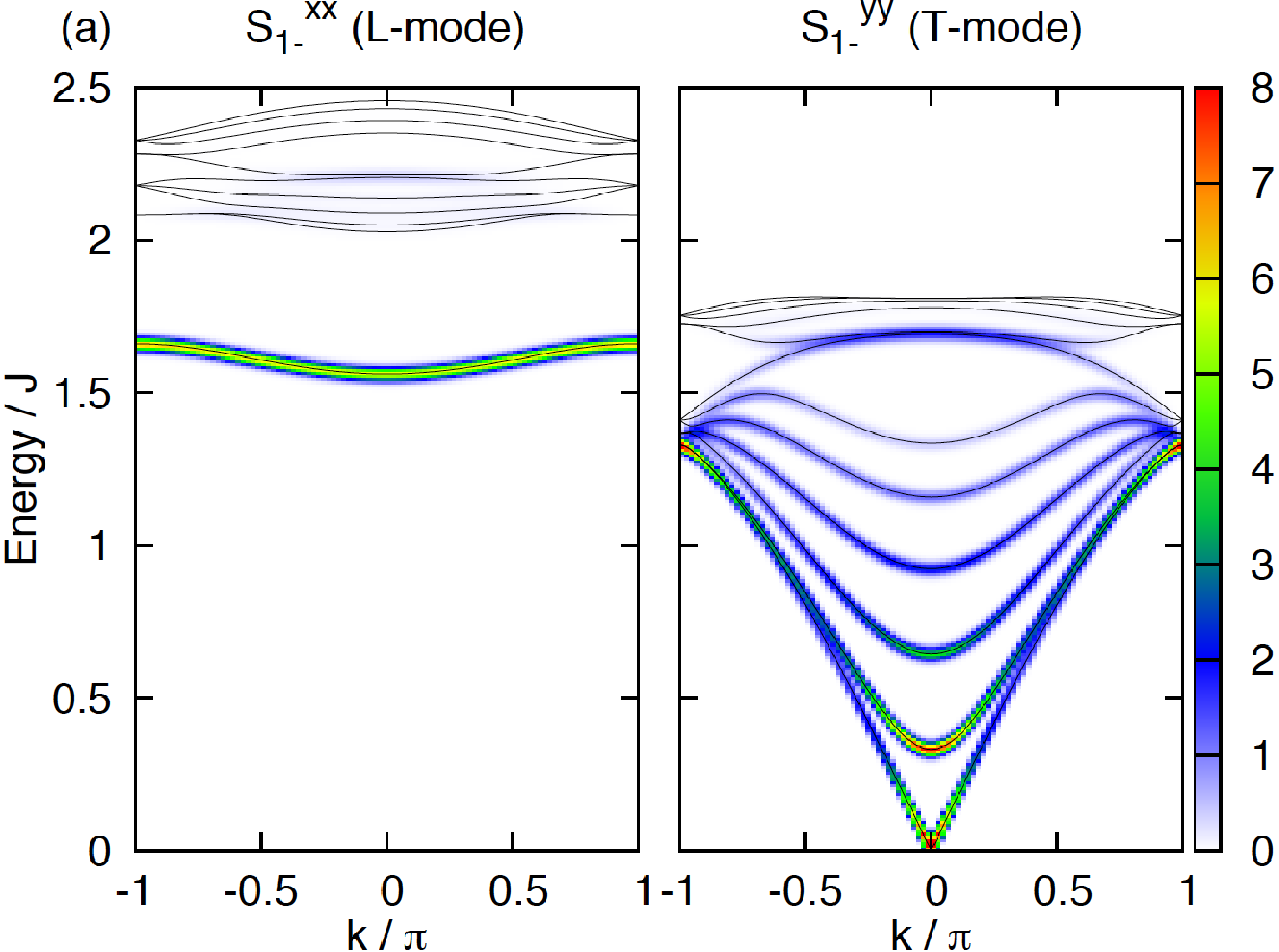}
\end{center}
\begin{center}
\includegraphics[width=7cm,clip]{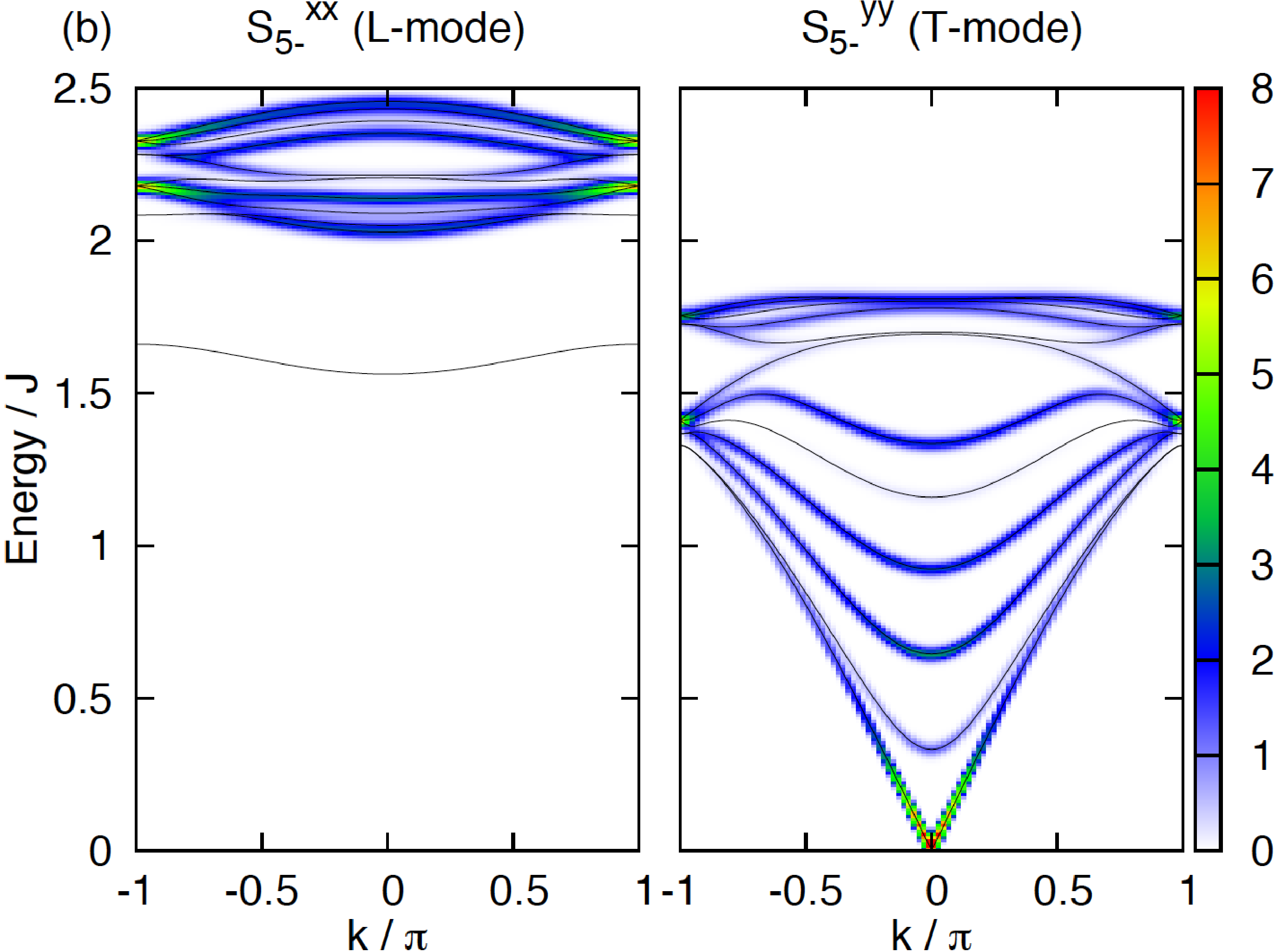}
\end{center}
\caption{
(Color online)
Dynamical spin correlation function in the ordered phase under $h=0$ for $J'=2.2J'_{\rm c}$.
The thin black line represents the magnon dispersion relation.
(a) For the edge site $(i_x=1)$.
(b) For the bulk site $(i_x=5)$.
}
\label{fig:sq-jp}
\end{figure}

To see the localized nature of the edge mode,
we next show the correlation function in Fig. \ref{fig:sq-jp} for $J'=2.2 J'_{\rm c}$.
We take the $x$-coordinate along the AF moment.
Since the $y$- and $z$-directions are equivalent for $h=0$, $S_{i_x-}^{yy}=S_{i_x-}^{zz}$ is satisfied.
In Fig. \ref{fig:sq-jp}(a), we can see a strong intensity on the lowest L-mode for the edge site.
This is because the potential for the magnon at the edge site is less than that for the bulk site.
In the case of the T-mode, the lowest mode has a stronger intensity at the edge site,
especially in the region of $|k|/\pi>0.5$.
This means that the edge mode moves from the flat position to the lowest position
when the phase becomes deep in the ordered phase.
This also holds in the field-induced ordered phase.
Note that the lowest mode in Fig. \ref{fig:energy-h}(f) also has the localized character.
In the next section, we discuss this point in connection with the spin system on a monolayer honeycomb lattice.

\section{Monolayer Spin System}
\label{sec:edge}

To understand the edge mode in the ordered phase of the dimer system,
we study $S=1/2$ spin systems on a monolayer honeycomb lattice with a zigzag edge.
In the case of a ferromagnet, a magnon moves by the hopping process,
while it moves by the pair creation and annihilation process in the AF case.
As discussed in Sect. \ref{sec:analytic-dimer}, the dimer system has both processes.
In the following subsections, we study how the edge mode is affected by the two processes
in the presence of an ordered moment.

\subsection{Ferromagnetic case}
\label{sec:F}

In the ferromagnetic case, the excitation moves by the hopping process.
Therefore, a particle and a hole are not coupled here.
In this case, Eq. (\ref{eqn:abcd}) reduces to
\begin{align}
\frac{3}{2}Ja_{n} + g b_{n-1} + f b_{n} &= E a_{n}, \cr
\frac{3}{2}Jb_{n} + f a_{n} + g a_{n+1} &= E b_{n},
\label{eqn:abcd-F}
\end{align}
with
\begin{align}
f = - J\cos\left(\frac{k}{2}\right),~~~
g = - \frac{1}{2}J.
\end{align}
Here, $J(>0)$ represents the strength of the ferromagnetic exchange interaction.
Note that Eq. (\ref{eqn:abcd-F}) is the same as that for graphene
and that the energy of the bulk excitation is simply shifted by $\frac{3}{2}J$ from that for graphene.
At the edge site, we write the boundary condition as
\begin{align}
\left( \frac{3}{2} + \delta \right) J a_{1} + f b_{1} = E a_{1}.
\label{eqn:boundary-F}
\end{align}
Here, $\delta$ represents the deviation of the potential term from the bulk value.
In the ferromagnetic ordered state with a full moment, $\delta=-\frac{1}{2}$,
while $\delta=0$ represents a uniform potential.
The latter situation can be realized by applying a local magnetic field at the edge site.
To understand the effect of a nonuniform potential, we study a general value of $\delta$ here.

In the same manner as in Sect. \ref{sec:analytic-dimer}, the equation for the bulk solution is given by
\begin{align}
\left(
\begin{matrix}
\frac{3}{2}J-E & f + \frac{g}{z} \cr
f+zg & \frac{3}{2}J-E
\end{matrix}
\right)
\left(
\begin{matrix}
a_0 \cr
b_0
\end{matrix}
\right) = 0.
\label{eqn:F-bulk}
\end{align}
Equation (\ref{eqn:F-bulk}) leads to the following boundary energies for the bulk excitations:
\begin{align}
&E_{k,1\pm}=J \left[ \frac{3}{2} + \left| \frac{1}{2}\pm\cos\left(\frac{k}{2}\right) \right| \right], \cr
&E_{k,2\pm}=J \left[ \frac{3}{2} - \left| \frac{1}{2}\pm\cos\left(\frac{k}{2}\right) \right| \right].
\label{eqn:F-boundary}
\end{align}

For the edge solution, we compare Eqs. (\ref{eqn:boundary-F}) and (\ref{eqn:abcd-F}) for $n=1$.
We notice that the following relation must be satisfied:
\begin{align}
-\delta J a_1 + g b_0 = 0.
\label{eqn:condition-F-2}
\end{align}
Equation (\ref{eqn:condition-F-2}) leads to
\begin{align}
z = -\frac{b_0}{2\delta a_0}.
\label{eqn:z-F}
\end{align}
Substituting Eq. (\ref{eqn:z-F}) into Eq. (\ref{eqn:F-bulk}), we obtain
\begin{align}
\left(
\begin{matrix}
\left( \frac{3}{2} + \delta \right) J - E & f \cr
f & \left( \frac{3}{2} + \frac{1}{4\delta} \right) J - E
\end{matrix}
\right)
\left(
\begin{matrix}
a_0 \cr
b_0
\end{matrix}
\right) = 0.
\label{eqn:F-edge}
\end{align}
The eigenvalue equation can be expressed analytically.
The condition for the edge mode is given by $|z|<1$.
For $\delta=-\frac{1}{2}$ and $\delta=0$, the energy eigenvalues are simplified to
\begin{align}
E_{k,{\rm edge}} =
\begin{cases}
J \left[ 1 \pm \cos\left(\frac{k}{2}\right) \right]~(\delta=-\frac{1}{2},~|z|=1) \cr
\frac{3}{2}J~~~~~~~~~~~~~~~~~~(\delta=0,~|k|>\frac{2\pi}{3})
\end{cases}.
\end{align}
For $\delta=-\frac{1}{2}$, the mode has a real $z$ value; however, it is a marginal state with $|z|=1$ for all $k$.
For $\delta=0$, we obtain the same bulk and flat edge excitation modes as those for graphene.

\subsection{$S=1/2$ AF system}

For the AF case, Eq. (\ref{eqn:abcd}) reduces to
\cite{You-2008}
\begin{align}
\frac{3}{2}Ja_{n} - g \bar{b}_{n-1} - f \bar{b}_{n} &= E a_{n}, \cr
-\frac{3}{2}J\bar{b}_{n} + f a_{n} + g a_{n+1} &= E \bar{b}_{n},
\label{eqn:abcd-AF}
\end{align}
with
\begin{align}
f=J\cos\left(\frac{k}{2}\right),~~~
g=\frac{1}{2}J.
\end{align}
At the edge site, we write
\begin{align}
\left( \frac{3}{2} + \delta \right) Ja_{1} - f \bar{b}_{1} &= E a_{1}.
\label{eqn:boundary-AF}
\end{align}
Here, $\delta$ represents the deviation of the potential term from the bulk value.
$\delta=-\frac{1}{2}$ represents the potential for the AF ordered state with a full moment,
while $\delta=0$ represents a uniform potential.

For the bulk solution, we obtain
\begin{align}
\left(
\begin{matrix}
\frac{3}{2}J-E & -f-\frac{g}{z} \cr
f+zg & -\frac{3}{2}J-E
\end{matrix}
\right)
\left(
\begin{matrix}
a_0 \cr
\bar{b}_0
\end{matrix}
\right) = 0.
\label{eqn:AF-bulk}
\end{align}
The boundary energies for the bulk excitations are expressed as
\cite{You-2008}
\begin{align}
E_{k\pm} = J \sqrt{ 1 + \sin^2\left(\frac{k}{2}\right)\pm\cos\left(\frac{k}{2}\right) }.
\label{eqn:AF-boundary}
\end{align}

Next, we study the edge solution.
Comparing Eqs. (\ref{eqn:boundary-AF}) and (\ref{eqn:abcd-AF}) for $n=1$,
we notice that the following relation must be satisfied:
\begin{align}
\delta J a_1 + g \bar{b}_0 = 0.
\label{eqn:condition-AF-2}
\end{align}
Equation (\ref{eqn:condition-AF-2}) leads to
\begin{align}
z = - \frac{\bar{b}_0}{2\delta a_0}.
\label{eqn:z-AF}
\end{align}
Substituting Eq. (\ref{eqn:z-AF}) into Eq. (\ref{eqn:AF-bulk}), we obtain
\begin{align}
\left(
\begin{matrix}
\left(\frac{3}{2}+\delta\right) J - E & -f \cr
f & - \left( \frac{3}{2} + \frac{1}{4\delta} \right) J - E
\end{matrix}
\right)
\left(
\begin{matrix}
a_0 \cr
\bar{b}_0
\end{matrix}
\right) = 0.
\label{eqn:AF-edge}
\end{align}
We can solve the eigenvalue equation analytically.
For $\delta=-\frac{1}{2}$ and $\delta=0$, the edge mode is expressed as
\cite{You-2008}
\begin{align}
E_{k,{\rm edge}} =
\begin{cases}
J \left| \sin\left(\frac{k}{2}\right) \right|~~~(\delta=-\frac{1}{2}) \cr
\frac{3}{2}J~~~~~~~~~~~~~~(\delta=0,~|k|>\frac{2\pi}{3})
\end{cases}.
\end{align}

In the following section, we discuss the edge state for a general $\delta$ value
in connection with that for the dimer in the ordered phase.

\section{Summary and Discussion}

We studied the edge magnon excitation in spin dimer systems on a bilayer honeycomb lattice with a zigzag edge
and found that an edge mode with a completely flat dispersion relation appears in the disordered phase.
In the ordered phase, the edge mode becomes dispersive and shifts to a low-energy region
with the development of an ordered moment.

\begin{figure}[t]
\begin{center}
\includegraphics[width=4cm,clip]{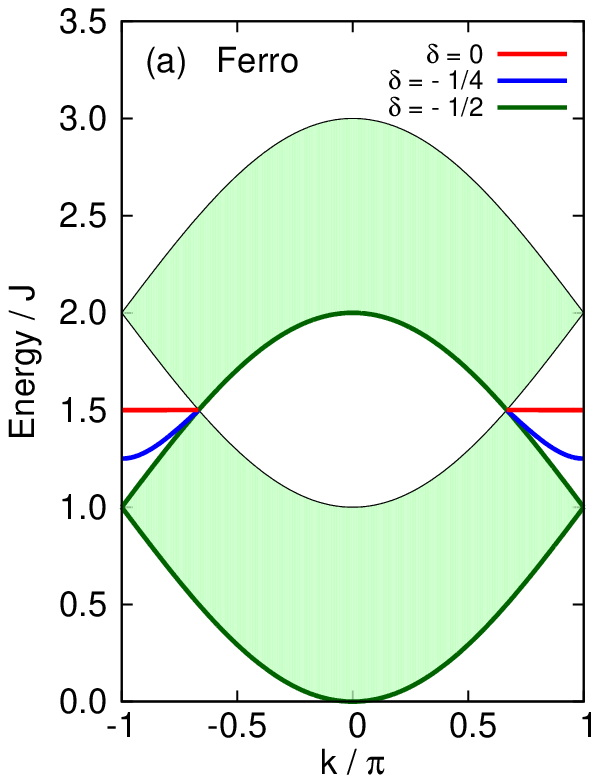}
\includegraphics[width=4cm,clip]{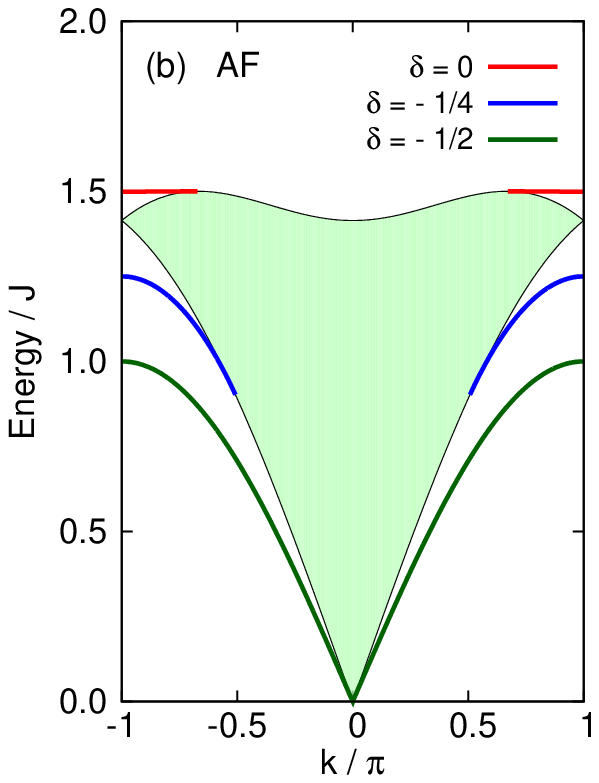}
\end{center}
\caption{
(Color online)
Energy region of the bulk magnon excitation and edge magnon modes for the $S=1/2$ spin system
on a monolayer honeycomb lattice with a zigzag edge.
The strength of the intersite interaction is represented by $J(>0)$.
(a) Ferromagnetic case.
The filled area represents the energy region for the bulk excitation.
The energy boundary for the bulk excitation is plotted with Eq. (\ref{eqn:F-boundary}).
The solid lines represent the edge magnon mode for various values of $\delta$ (see Sect. \ref{sec:edge}).
They are plotted by solving Eq. (\ref{eqn:F-edge}) under $|z|\le 1$.
(b) AF case.
The energy boundary for the bulk excitation is plotted with Eq. (\ref{eqn:AF-boundary}).
The edge mode is plotted by solving Eq. (\ref{eqn:AF-edge}) under $|z|\le 1$.
It continues until it meets the region for the bulk excitation.
For $\delta=-(\frac{3}{2}-\sqrt{2})\simeq -0.0856$, no edge mode appears.
\label{fig:energy-edge}
}
\end{figure}

\begin{table}[t]
\caption{
Summary of the edge mode on the honeycomb lattice with a zigzag edge in various substances.
In the `Substance' category, ferromagnet and antiferromagnet refer to monolayer systems.
In the `Potential' category, U and N represent uniform and nonuniform potentials, respectively.
They are related to the dispersion relation of the edge mode.
For U and N, the edge mode is flat and dispersive, respectively.
In the `Movement' category, hop and pair indicate that the excitation moves
via the hopping and pair creation and annihilation processes, respectively.
For a dimer system in the ordered phase with a large moment,
the pair term is the main contribution to the motion of the excited state.
In the `Edge mode' category, middle and below indicate that the edge mode
appears in the middle of and below the bulk excitations, respectively.
\cite{pair}
}
\begin{tabular}{lcll}
\hline
Substance          & Potential & Movement   &~~~~Edge mode \\
\hline
graphene           & U         & hop        & flat,~~~~~~~~~middle \\
ferromagnet        & N         & hop        & dispersive, middle   \\
dimer (disordered) & U         & hop + pair & flat,~~~~~~~~~middle \\
dimer (ordered)    & N         & pair       & dispersive, below    \\
antiferromagnet    & N         & pair       & dispersive, below    \\
\hline
\end{tabular}
\label{table:edge-mode}
\end{table}

The result in the ordered phase is related to the spin systems
on a monolayer honeycomb lattice studied in Sect. \ref{sec:edge}.
Figure \ref{fig:energy-edge} shows the edge mode of the $S=1/2$ spin system in both the ferromagnetic and AF cases,
where the magnon moves via the hopping and pair creation and annihilation processes, respectively.
In the ferromagnetic case [see Fig. \ref{fig:energy-edge}(a)],
a flat edge mode appears in the middle of the bulk excitations under a uniform potential ($\delta=0$).
When the potential becomes nonuniform ($\delta=-\frac{1}{4}$), the edge mode becomes dispersive.
As shown in Fig. \ref{fig:sq-1.1-hc}(a) for a spin dimer,
we can see similar behavior of the edge mode in the ordered phase.
In the case of the spin dimer, both processes contribute to the motion of the excited state.
The bulk excitations both below and above the edge mode are owing to the hopping term,
while their asymmetry is from the pair creation and annihilation term.
Thus, the dispersive edge mode in the vicinity of the quantum critical point shown in Fig. \ref{fig:sq-1.1-hc}(a)
originates from the nonuniform potential arising in the ordered phase.

When the phase is deep in the ordered phase, a large ordered moment appears,
where the singlet and triplet components become comparable in the ground state.
In this case, the lowest modes in Figs. \ref{fig:energy-h}(f) and \ref{fig:energy-jp-order}(f)
acquire the localized character.
Under a large ordered moment, the hopping term is strongly reduced,
as in Eq. (\ref{eqn:hop-pair}) with $u\simeq v$, for instance, in the interdimer interaction-induced ordered phase.
In addition, the potential term at the edge differs markedly from that at the bulk site.
We notice that these properties are similar to the AF monolayer system,
where the edge mode shifts to the low-energy region and is dispersive
when the potential becomes strongly nonuniform [see Fig. \ref{fig:energy-edge}(b)].
Thus, the edge mode in the deep ordered phase can be understood by the AF monolayer system.
Table \ref{table:edge-mode} summarizes the edge mode in various substances.

In spin dimer systems, the disordered phase is realized for a large intradimer interaction.
This is peculiar to spin dimer systems and gives rise to the magnon mode with a completely flat dispersion relation.
For integer spin systems on a monolayer system, strong easy-plane anisotropy also stabilizes the disordered phase
with a singlet-doublet spin multiplet configuration.
Since the multiplet plays the role of a dimer, the flat edge magnon mode also appears in this case.
Searching for the edge magnon excitation is an interesting subject to study.
Note that the magnon dispersion relation of the edge mode is easily controllable
by applying a magnetic field or pressure in the ordered phase, differing from the case of graphene.
When the edge magnon is excited selectively by tuning the excitation energy,
a magnetic moment appears dynamically only in the vicinity of the edge.
Thus, the edge magnon can be used to control the edge magnetic moment.
The existence of the edge magnon mode has potential application to future spintronics devices.

\acknowledgments

This work was supported by JSPS KAKENHI Grant Number 26400332.

\appendix
\setcounter{equation}{0}

\section{Bond Operator Theory for Infinite Bilayer Honeycomb Lattice}
\label{appendix-bulk}

\subsection{Disordered phase}

For spin dimer systems on an infinite lattice, magnetic excitations are expressed analytically in the disordered phase.
The model Hamiltonian for an infinite honeycomb lattice is given by Eq. (\ref{eqn:H-dimer}).
The energy eigenstates of an isolated dimer are expressed as
\begin{align}
&|{\rm s}\rangle = \frac{1}{\sqrt{2}} \left( |\uparrow\downarrow\rangle - |\downarrow\uparrow\rangle \right), \cr
&|{\rm 0}\rangle = \frac{1}{\sqrt{2}} \left( |\uparrow\downarrow\rangle + |\downarrow\uparrow\rangle \right), \cr
&|{\rm +}\rangle = |\uparrow\uparrow\rangle,~~~
|{\rm -}\rangle = |\downarrow\downarrow\rangle.
\label{eqn:s-t}
\end{align}
Here, $|\sigma\sigma'\rangle$ represents a spin state of a dimer.
$|{\rm s}\rangle$ is a singlet state, while $|m\rangle~(m=+,0,-)$ is a triplet state.
We introduce the following bosonic creation operators for the four states at each dimer site:
\cite{Sachdev-1990}
\begin{align}
&|{\rm s}_i\rangle = s_i^\dagger |{\rm vac}\rangle,~~~
|m_i\rangle = t_{im}^\dagger |{\rm vac}\rangle~~~(m=+,0,-).
\end{align}
Here, $|{\rm vac}\rangle$ represents the vacuum state.
$s_i^\dagger$ and $t_{i m}^\dagger$ are creation operators for the singlet and triplet states at the $i$th site, respectively.
They are subjected to the following local constraint:
\begin{align}
s_i^\dagger s_i + \sum_{m=+,0,-} t_{im}^\dagger t_{im} = 1.
\label{eqn:constraint}
\end{align}
The Hamiltonian for a dimer is then expressed as
\begin{align}
J \bS_{i{\rm l}} \cdot \bS_{i{\rm r}} = -\frac{3}{4}J s_i^\dagger s_i + \frac{1}{4}J \sum_{m=+,0,-} t_{im}^\dagger t_{im}.
\end{align}
The spin operators are also expressed in terms of the bosons as
\begin{align}
&S_{i{\rm l}}^x = \frac{1}{2\sqrt{2}}
\left( -t_{i+}^\dagger s_i + s_i^\dagger t_{i-} + t_{i+}^\dagger t_{i0} + t_{i0}^\dagger t_{i-} \right) + {\rm h.c.}, \cr
&S_{{i\rm r}}^x = \frac{1}{2\sqrt{2}}
\left( t_{i+}^\dagger s_i - s_i^\dagger t_{i-} + t_{i+}^\dagger t_{i0} + t_{i0}^\dagger t_{i-} \right) + {\rm h.c.}, \cr
&S_{i{\rm l}}^y = \frac{-i}{2\sqrt{2}}
\left( -t_{i+}^\dagger s_i + s_i^\dagger t_{i-} + t_{i+}^\dagger t_{i0} + t_{i0}^\dagger t_{i-} \right) + {\rm h.c.}, \cr
&S_{i{\rm r}}^y = \frac{-i}{2\sqrt{2}}
\left( t_{i+}^\dagger s_i - s_i^\dagger t_{i-} + t_{i+}^\dagger t_{i0} + t_{i0}^\dagger t_{i-} \right) + {\rm h.c.}, \cr
&S_{i{\rm l}}^z = \frac{1}{2}
\left( t_{i0}^\dagger s_i + s_i^\dagger t_{i0} + t_{i+}^\dagger t_{i+} - t_{i-}^\dagger t_{i-} \right), \cr
&S_{i{\rm r}}^z = \frac{1}{2}
\left( -t_{i0}^\dagger s_i - s_i^\dagger t_{i0} + t_{i+}^\dagger t_{i+} - t_{i-}^\dagger t_{i-} \right).
\label{eqn:spin-operator}
\end{align}

In the disordered phase, the mean-field ground state is the singlet state at each site.
This is described by Bose-Einstein condensation of the singlet boson
and the singlet operators are replaced as $s_{i}\rightarrow 1$ and $s_{i}^\dagger\rightarrow 1$.
The intradimer part of the Hamiltonian is then expressed as
\begin{align}
\H_{\rm intra} = \sum_i \sum_{m=+,0,-} (J-mh) t_{im}^\dagger t_{im} + E_0.
\label{eqn:H-intra}
\end{align}
Here, $E_0 = - \frac{3}{4}JN$
and it represents the energy of the mean-field ground state with $N$ the number of dimer sites.
$h(=g\mu_{\rm B} H_z)$ represents the effective magnetic field applied along the $z$-direction.
For the interdimer part of the Hamiltonian, the spin operators are expressed up to the first order of the triplet bosons as
\begin{align}
&S_{i{\rm l}}^x = -S_{i{\rm r}}^x = \frac{1}{2\sqrt{2}}
\left( -t_{i+}^\dagger + t_{i-} - t_{i+} + t_{i-}^\dagger \right), \cr
&S_{i{\rm l}}^y = -S_{i{\rm r}}^y = \frac{-i}{2\sqrt{2}}
\left( -t_{i+}^\dagger + t_{i-} + t_{i+} - t_{i-}^\dagger \right), \cr
&S_{i{\rm l}}^z = -S_{i{\rm r}}^z = \frac{1}{2}
\left( t_{i0}^\dagger + t_{i0} \right).
\label{eqn:S-operator}
\end{align}
Substituting Eqs. (\ref{eqn:H-intra}) and (\ref{eqn:S-operator}) into Eq. (\ref{eqn:H-dimer}), we obtain
\begin{align}
&\H = \H_0 + \H_\pm + E_0,
\end{align}
where
\begin{align}
&\H_0 = \sum_i J t_{i0}^\dagger t_{i0}
+ \sum_{\langle ij\rangle} \frac{1}{2} J'
\left( t_{i0}^\dagger t_{j0} + t_{i0}^\dagger t_{j0}^\dagger + {\rm h.c.} \right), \cr
&\H_\pm = \sum_i \sum_{m=\pm} (J-mh) t_{im}^\dagger t_{im} \cr
&~~~
+ \sum_{\langle ij\rangle} \frac{1}{2} J'
  \sum_{m=\pm} \left( t_{im}^\dagger t_{jm} - t_{im}^\dagger t_{j,-m}^\dagger + {\rm h.c.} \right).
\label{eqn:Hij}
\end{align}
On each sublattice, we introduce the following Fourier transformations for the operator:
\begin{align}
&t_{im} = \frac{1}{\sqrt{\frac{N}{2}}} \sum_\bk e^{i\bk\cdot\bm{r}_i} a_{\bk m}~~~(i={\rm A}), \cr
&t_{im} = \frac{1}{\sqrt{\frac{N}{2}}} \sum_\bk e^{i\bk\cdot\bm{r}_i} b_{\bk m}~~~(i={\rm B}).
\label{eqn:Fourier}
\end{align}
Here $a_{\bk m}$ and $b_{\bk m}$ are Fourier transformed operators
for the A and B sublattices shown in Fig. \ref{fig:honeycomb}, respectively.
$\bm{r}_i$ represents the position of the $i$th dimer.
Substituting Eq. (\ref{eqn:Fourier}) into Eq. (\ref{eqn:Hij}), we obtain
\begin{align}
&\H_0 = \sum_\bk \left\{ J \left( a_{\bk 0}^\dagger a_{\bk 0} + b_{\bk 0}^\dagger b_{\bk 0} \right) \right. \cr
&~~~~~~\left.
+ \left[ \gamma_\bk \left( a_{\bk 0}^\dagger b_{\bk 0} + a_{\bk 0}^\dagger b_{\bk 0}^\dagger \right)
         + {\rm h.c.} \right] \right\}, \cr
&\H_\pm = \sum_\bk \sum_{m=\pm} \left\{
          (J-mh) \left( a_{\bk m}^\dagger a_{\bk m} + b_{\bk m}^\dagger b_{\bk m} \right) \right. \cr
&~~~~~~\left.
+ \left[ \gamma_\bk \left( a_{\bk m}^\dagger b_{\bk m} - a_{\bk m}^\dagger b_{\bk,-m}^\dagger \right)
         + {\rm h.c.} \right] \right\},
\label{eqn:Hk}
\end{align}
where $\gamma_\bk$ is defined by
\begin{align}
\gamma_\bk = \frac{1}{2} J' \left( e^{i\bk\cdot\bd_1} + e^{i\bk\cdot\bd_2} + e^{i\bk\cdot\bd_3} \right)
= |\gamma_\bk| e^{i\phi_\bk}.
\label{eqn:gamma-k}
\end{align}
Here, $\phi_\bk$ represents the phase factor of $\gamma_\bk$,
and $\bd_n~(n=1,2,3)$ is given in the caption of Fig. \ref{fig:honeycomb}.
For later convenience, we introduce the following bosons:
\cite{Weihong-1991}
\begin{align}
&a_{\bk m} = \frac{1}{\sqrt{2}} \left( c_{\bk m +} + c_{\bk m -} \right) e^{i\phi_\bk},~~~(m=+,0,-) \cr
&b_{\bk m} = \frac{1}{\sqrt{2}} \left( c_{\bk m +} - c_{\bk m -} \right).
\label{eqn:transform}
\end{align}
Substituting Eq. (\ref{eqn:transform}) into Eq. (\ref{eqn:Hk}), we obtain
\begin{align}
&\H_0 = \sum_\bk \sum_{\tau=\pm} \left[
\epsilon_{\bk 0 \tau} c_{\bk 0 \tau}^\dagger c_{\bk 0 \tau}
+ \tau \frac{1}{2} \Delta_\bk \left( c_{\bk 0 \tau}^\dagger c_{\bk 0 \tau}^\dagger + {\rm h.c.} \right) \right], \cr
&\H_\pm = \sum_\bk \sum_{\tau=\pm} \sum_{m=\pm} \left[
\epsilon_{\bk m \tau} c_{\bk m \tau}^\dagger c_{\bk m \tau} \right. \cr
&~~~~~~~~~~~~~~~~~~\left.
- \tau \frac{1}{2} \Delta_\bk \left( c_{\bk m \tau}^\dagger c_{\bk,-m \tau}^\dagger + {\rm h.c.} \right) \right],
\label{eqn:H-c-tau}
\end{align}
where
\begin{align}
&\epsilon_{\bk m \tau} = J -mh + \tau |\gamma_\bk|,~~~(m=+,0,-) \cr
&\Delta_\bk = |\gamma_\bk|.
\end{align}
To diagonalize the Hamiltonian in Eq. (\ref{eqn:H-c-tau}), we introduce the following Bogoliubov transformation:
\begin{align}
&c_{\bk 0 \tau} = u_{\bk\tau} \alpha_{\bk 0 \tau} - v_{\bk\tau} \alpha_{-\bk 0 \tau}^\dagger, \\
&c_{\bk m \tau} = u_{\bk\tau} \alpha_{\bk m \tau} + v_{\bk\tau} \alpha_{-\bk,-m \tau}^\dagger.~~~(m=\pm) \nonumber
\end{align}
The Hamiltonian is then diagonalized as
\begin{align}
\H &= \H_0 + \H_\pm + E_0 \cr
&= \sum_\bk \sum_{\tau=\pm} \sum_{m=+,0,-} \left\{
E_{\bk m\tau} \alpha_{\bk m \tau}^\dagger \alpha_{\bk m \tau} \right. \cr
&~~~~~~~~~~~~~~~\left.
- \left[ \frac{3}{4}J + \frac{1}{2} \left( \epsilon_{\bk 0 \tau} - E_{\bk 0\tau} \right) \right] \right\},
\label{eqn:H-final}
\end{align}
where
\begin{align}
&E_{\bk m\tau} = \sqrt{ \epsilon_{\bk 0 \tau}^2 - \Delta_\bk^2 } - mh, \label{eqn:H-final-2} \\
&u_{\bk\tau} = \sqrt{ \frac{1}{2} \left( \frac{\epsilon_{\bk 0 \tau}}{E_{\bk 0\tau}} + 1 \right) },~~~
v_{\bk\tau} = \sqrt{ \frac{1}{2} \left( \frac{\epsilon_{\bk 0 \tau}}{E_{\bk 0\tau}} - 1 \right) }. \nonumber
\end{align}
In Eq. (\ref{eqn:H-final}), there are constant terms.
The first term is the energy of the mean-field ground state,
while the second term is its quantum correction from the spin-wave excitations.
The $\alpha_{\bk m \tau}$ boson describes the magnon excitation.
There are six excitation modes in the first Brillouin zone, replicating the two sublattices in the unit cell.
$E_{\bk m \tau}$ represents the magnon dispersion relation.
The excitations are threefold degenerate with respect to $m=+,0,-$ under $h=0$.
For a finite magnetic field, they are split into three branches by the Zeeman effect.

In the disordered phase, the magnons have a finite excitation gap.
The lowest excitation is at $\bk=(0,0)$ ($\Gamma$ point) for $\tau=-$.
Since the excitation gap closes at the critical field $H_{\rm c1}$, it is determined as
\begin{align}
h_{\rm c1} = g\mu_{\rm B} H_{\rm c1} = E_{\bk=0,m=+,\tau=-} = \sqrt{ J^2 - 3JJ'}.
\label{eqn:hc1}
\end{align}
In the absence of an external field, the excitation gap also closes when the interdimer interaction $J'$ is increased.
The critical value of the interdimer interaction is determined as
\begin{align}
J_{\rm c}' = \frac{1}{3}J.
\label{eqn:Jpc}
\end{align}

In the saturated phase, the mean-field ground state is the $|\uparrow\uparrow\rangle$ triplet state,
where the magnetic excitation is also given analytically.
In the same way as in the disordered phase, the saturation field is given by
\begin{align}
h_{\rm c2} = g\mu_{\rm B} H_{\rm c2} = J + 3J'.
\label{eqn:hc2}
\end{align}

\subsection{Ordered phase under $h=0$}

In the absence of a magnetic field, the formulation of the excitation mode becomes simple even in the ordered phase.
Under $h=0$, it is convenient to introduce the following triplet states:
\cite{Sachdev-1990}
\begin{align}
&|x\rangle = \frac{1}{\sqrt{2}} \left( - |\uparrow\uparrow\rangle + |\downarrow\downarrow\rangle \right), \cr
&|y\rangle = \frac{i}{\sqrt{2}} \left( |\uparrow\uparrow\rangle + |\downarrow\downarrow\rangle \right), \cr
&|z\rangle = \frac{1}{\sqrt{2}} \left( |\uparrow\downarrow\rangle + |\downarrow\uparrow\rangle \right).
\label{eqn:txyz}
\end{align}
The spin operators are then expressed as
\cite{Sachdev-1990}
\begin{align}
&S_{i{\rm l}}^\alpha = \frac{1}{2} \left( t_{i\alpha}^\dagger s_i + s_i^\dagger t_{i\alpha}
                                        - i \epsilon_{\alpha\beta\gamma} t_{i\beta}^\dagger t_{i\gamma} \right), \cr
&S_{i{\rm r}}^\alpha = \frac{1}{2} \left( - t_{i\alpha}^\dagger s_i - s_i^\dagger t_{i\alpha}
                                          - i \epsilon_{\alpha\beta\gamma} t_{i\beta}^\dagger t_{i\gamma} \right).
\end{align}
Here, $\epsilon_{\alpha\beta\gamma}$ represents the antisymmetric tensor
and $\alpha$, $\beta$, and $\gamma$ take $x$, $y$, and $z$.
As in the disordered phase, we introduce the $t_{i\alpha}^\dagger$ boson for the $|\alpha\rangle$ state.

We take the $z$-axis along the ordered moment here.
On the up (A) site, we introduce the following transformations for the operators:
\cite{Sommer-2001,Matsumoto-2002,Matsumoto-2004}
\begin{align}
&s_{i} = u a_{i0} - v a_{iL},~~~
t_{iz} = u a_{iL} + v a_{i0}, \cr
&t_{i\alpha} = a_{i\alpha}~~~(\alpha=x,y).
\label{eqn:transform-a}
\end{align}
Here, $a_{i0}$ is a boson for the mean-field ground state.
$a_{iL}$ describes the L-mode, while $a_{ix}$ and $a_{iy}$ describe the T-mode.
$u$ and $v$ are real coefficients satisfying
\begin{align}
u^2 + v^2 = 1.
\end{align}
They are determined to minimize the energy of the mean-field ground state.
At each site, the bosons are subjected to the following local constraint:
\begin{align}
a_{i0}^\dagger a_{i0} + \sum_{m=L,x,y} a_{im}^\dagger a_{im} = 1.
\label{eqn:constraint-J'}
\end{align}
Using this constraint, we replace the $a_{i0}$ boson as
\cite{Sommer-2001,Matsumoto-2002,Matsumoto-2004}
\begin{align}
&a_{i0}^\dagger a_{i0} \rightarrow 1 - \sum_{m=L,x,y} a_{im}^\dagger a_{im}, \cr
&a_{i0} \rightarrow \left( 1 - \sum_{m=L,x,y} a_{im}^\dagger a_{im} \right)^{1/2}, \cr
&a_{i0}^\dagger \rightarrow \left( 1 - \sum_{m=L,x,y} a_{im}^\dagger a_{im} \right)^{1/2}.
\label{eqn:replace-2}
\end{align}
Up to the quadratic order of bosons, the spin operators are expressed as
\begin{align}
&S_{il}^x = \frac{1}{2} \left[ u ( a_{ix}^\dagger + a_{ix} ) - iv ( a_{iy}^\dagger - a_{iy} ) \right], \cr
&S_{ir}^x = \frac{1}{2} \left[ - u ( a_{ix}^\dagger + a_{ix} ) - iv ( a_{iy}^\dagger - a_{iy} ) \right], \cr
&S_{il}^y = \frac{1}{2} \left[ u ( a_{iy}^\dagger + a_{iy} ) + iv ( a_{ix}^\dagger - a_{ix} ) \right], \cr
&S_{ir}^y = \frac{1}{2} \left[ - u ( a_{iy}^\dagger + a_{iy} ) + iv ( a_{ix}^\dagger - a_{ix} ) \right], \cr
&S_{il}^z = \frac{1}{2} \left[ (u^2-v^2) ( a_{iL}^\dagger + a_{iL} )
                           - i ( a_{ix}^\dagger a_{iy} - a_{iy}^\dagger a_{ix} ) \right. \cr
&\left.~~~~~~~~
   + 2uv \left( 1 - 2a_{iL}^\dagger a_{iL} - \sum_{m=x,y} a_{im}^\dagger a_{im} \right) \right], \cr
&S_{ir}^z = \frac{1}{2} \left[ - (u^2-v^2) ( a_{iL}^\dagger + a_{iL} )
                             - i ( a_{ix}^\dagger a_{iy} - a_{iy}^\dagger a_{ix} ) \right. \cr
&\left.~~~~~~~~
   - 2uv \left( 1 - 2a_{iL}^\dagger a_{iL} - \sum_{m=x,y} a_{im}^\dagger a_{im} \right) \right].
\label{eqn:spin-a}
\end{align}
Note that the c-number terms in $S_{il}^z$ and $S_{ir}^z$ are $uv$ and $-uv$, respectively.
They represent the ordered moment along the $z$-direction on the left and right sides of a dimer, respectively.
The intradimer interaction is also expressed as
\begin{align}
&J \bm{S}_{il}\cdot\bm{S}_{ir} = - \frac{3}{4} J s_{i}^\dagger s_{i} + \frac{1}{4} J \sum_{m=x,y,z} t_{im}^\dagger t_{im} \cr
&= - \frac{3}{4} J u^2 + \frac{1}{4} J v^2 + J uv ( a_{iL}^\dagger + a_{iL} )
\label{eqn:dimer-a} \\
&~~~
+ J (u^2-v^2) a_{iL}^\dagger a_{iL}
+ J u^2 ( a_{ix}^\dagger a_{ix} + a_{iy}^\dagger a_{iy} ).
\nonumber
\end{align}

Next, we consider the down (B) site.
For the B site, we introduce the following transforms:
\begin{align}
&s_{i} = u b_{i0} + v b_{iL},~~~
t_{iz} = u b_{iL} - v b_{i0}, \cr
&t_{i\alpha} = b_{i\alpha}~~~(\alpha=x,y).
\label{eqn:transform-b}
\end{align}
Here, $b_{i m}$ $(m=0,L,x,y)$ is the boson for the B site.
Comparing Eqs. (\ref{eqn:transform-a}) and (\ref{eqn:transform-b}),
we notice that the sign of $v$ is reversed for the B site.
This means that the ordered moment is staggered on the A and B sites.
The spin operators for the B site and the intradimer interaction are then expressed in the same way
as in Eqs. (\ref{eqn:spin-a}) and (\ref{eqn:dimer-a})
by replacing $v\rightarrow -v$ and $a_{i m}\rightarrow b_{i m}$.

Substituting Eqs. (\ref{eqn:spin-a}) and (\ref{eqn:dimer-a})
and the corresponding equations for the B site into Eq. (\ref{eqn:H-dimer}),
we obtain the following form of the Hamiltonian up to the second order of the Bose operators:
\begin{align}
\H = E_0 + \sum_{m=L,x,y} \H_m.
\label{eqn:H-order}
\end{align}
Here, $E_0$ is the energy of the mean-field ground state.
It is given by
\begin{align}
\frac{E_0}{N} &= \left( - \frac{3}{4}u^2 + \frac{1}{4}v^2 \right) J - 3(uv)^2 J' \cr
&= (J - 3J')v^2 + 3J' v^4 - \frac{3}{4}J.
\label{eqn:e0}
\end{align}
Note that $E_0$ corresponds to the Higgs potential.
Since we choose the $z$-axis along the ordered moment, the phase factor has been eliminated here.
The ordered phase is realized for a finite $v$.
The sign change of the coefficient of the $v^2$ term determines the critical value of $J'$.
It is given by $J'_{\rm c}=\frac{1}{3}J$, identically to Eq. (\ref{eqn:Jpc}).
For $J'\ge J'_{\rm c}$, $E_0$ is minimized with the following $u$ and $v$:
\begin{align}
u = \sqrt{ \frac{1}{2} \left( 1 + \frac{J'_{\rm c}}{J'} \right) },~~~
v = \sqrt{ \frac{1}{2} \left( 1 - \frac{J'_{\rm c}}{J'} \right) }.
\label{eqn:u-v}
\end{align}
For $J'\le J'_{\rm c}$, $E_0$ is minimized with $u=1$ and $v=0$.
The Hamiltonian $\H_m$ in Eq. (\ref{eqn:H-order}) is expressed in the following form:
\begin{align}
&\H_m = \sum_\bk \left[
e_m ( a_{\bk m}^\dagger a_{\bk m} + b_{\bk m}^\dagger b_{\bk m} ) \right. \cr
&~~~\left.
+ ( \epsilon_{\bk m} a_{\bk m}^\dagger b_{\bk m} + \Delta_{\bk m} a_{\bk m}^\dagger b_{-\bk m}^\dagger + {\rm h.c.} )
\right].
\label{eqn:Hm-jp}
\end{align}
Here, $a_{\bk m}$ and $b_{\bk m}$ are the Fourier transformed operators of $a_{i m}$ and $b_{i m}$, respectively.
Note that the first-order term in the Bose operator vanishes in the Hamiltonian by means of Eq. (\ref{eqn:u-v}).
The coefficients in the Hamiltonian are given by
\begin{align}
&e_L = (u^2-v^2)J + 12(uv)^2 J', \cr
&\epsilon_{\bk L} = (u^2-v^2)^2 \gamma_\bk,~~~
\Delta_{\bk L} = (u^2-v^2)^2 \gamma_\bk, \cr
&e_x = e_y = u^2 J + 6(uv)^2 J', \label{eqn:hop-pair} \\
&\epsilon_{\bk x} = \epsilon_{\bk y} = (u^2-v^2) \gamma_\bk,~~~
\Delta_{\bk x} = \Delta_{\bk y} = \gamma_\bk,
\nonumber
\end{align}
where $\gamma_\bk$ is defined by Eq. (\ref{eqn:gamma-k}).
$\epsilon_{\bk L}$ represents the hopping term,
while $\Delta_{\bk L}$ is the pair creation and annihilation term for the L-mode.
Similarly, $\epsilon_{\bk x}$, $\epsilon_{\bk y}$, $\Delta_{\bk x}$, and $\Delta_{\bk y}$ are the terms for the T-mode.
In the disordered phase, $u=1$ and $v=0$.
In this case, the hopping ($\epsilon_{\bk m}$) and pair ($\Delta_{\bk m}$) terms are equivalent.
In the deep ordered phase, where $u\simeq v$, the hopping term is greatly reduced.
This is owing to the fact that the singlet and triplet components in the ground state cancel the hopping.
Thus, the dispersive excitation mode is realized mainly by the pair term for the T-mode.
In contrast, the pair term ($\Delta_{\bk L}$) is also reduced for the L-mode.
Therefore, the L-mode becomes localized in the deep ordered phase.

The Hamiltonian $\H_m$ in Eq. (\ref{eqn:Hm-jp}) has the same form as $\H_0$ in Eq. (\ref{eqn:Hk}).
The excitation energy for the $m(=L,x,y)$ mode is then given by
\begin{align}
E_{\bk m \tau} = \sqrt{ \epsilon_{\bk m \tau}^2 - \Delta_{\bk m \tau}^2 }~~~(\tau=\pm),
\label{eqn:E-order}
\end{align}
where
\begin{align}
\epsilon_{\bk m \tau} = e_m + \tau |\epsilon_{\bk m}|,~~~
\Delta_{\bk m \tau} = \tau |\Delta_{\bk m}|.
\end{align}
$E_{\bk L\tau}$ is the excitation energy for the L-mode, while $E_{\bk x\tau}$ and $E_{\bk y\tau}$ are those for the T-mode.
In the disordered phase ($u=1$ and $v=0$),
Eq. (\ref{eqn:E-order}) reduces to Eq. (\ref{eqn:H-final-2}) when $h=0$.

\section{Specific Forms of $\Lambda_{mn}^{i_x}$ and $\gamma_k^{i_x j_x}$}
\label{sec:appendix-matrix}

\subsection{$\Lambda_{mn}^{i_x}$}

At the edge sites $(i_x=1, N_x)$, $\Lambda_{mn}^{i_x}$ is given by
\begin{align}
&\Lambda_{mn}^1 = \H_{mn}^{0}(1) - \H_{00}^{0}(1)\delta_{mn} \\
&+ J' \sum_\gamma \left[ \bS_{mn}(1,\gamma) - \bM_{1\gamma} \delta_{mn} \right] \cdot 2\bM_{2\gamma}, \cr
&\Lambda_{mn}^{N_x} = \H_{mn}^{0}(N_x) - \H_{00}^{0}(N_x)\delta_{mn} \\
&+ J' \sum_\gamma \left[ \bS_{mn}(N_x,\gamma) - \bM_{N_x\gamma} \delta_{mn} \right] \cdot 2\bM_{N_x-1\gamma}. \nonumber
\end{align}
For odd $i_x$ sites, it is given by
\begin{align}
&\Lambda_{mn}^{i_x} = \H_{mn}^{0}(i_x) - \H_{00}^{0}(i_x)\delta_{mn} \\
&+ J' \sum_\gamma \left[ \bS_{mn}(i_x,\gamma) - \bM_{i_x\gamma} \delta_{mn} \right] \cdot
\left[ 2\bM_{i_x+1\gamma} + \bM_{i_x-1\gamma} \right]. \nonumber
\end{align}
For even $i_x$ sites, it is given by
\begin{align}
&\Lambda_{mn}^{i_x} = \H_{mn}^{0}(i_x) - \H_{00}^{0}(i_x)\delta_{mn} \\
&+ J' \sum_\gamma \left[ \bS_{mn}(i_x,\gamma) - \bM_{i_x\gamma} \delta_{mn} \right] \cdot
\left[ 2\bM_{i_x-1\gamma} + \bM_{i_x+1\gamma} \right]. \nonumber
\end{align}

\subsection{$\gamma_k^{i_x j_x}$}

$\gamma_k^{i_x j_x}$ is given by
\begin{align}
&\gamma_k^{i_x+1,i_x} = \gamma_k^{i_x,i_x+1} = 2\cos\left(\frac{k}{2}\right)~~~({\rm for~odd}~i_x), \cr
&\gamma_k^{i_x+1,i_x} = \gamma_k^{i_x,i_x+1} = 1~~~~~~~~~~~~~~~~({\rm for~even}~i_x).
\end{align}

\section{Dynamical Spin Correlation Function}
\label{appendix-correlation}

First, we introduce the following Fourier transformation for the spin operator along the $y$-direction:
\begin{align}
S_{i_x\pm}^\alpha(q) = \frac{1}{\sqrt{N_y}} \sum_{i_y} e^{-i q i_y} S_{i\pm}^\alpha~~~(\alpha=x,y,z),
\label{eqn:Fourier-spin}
\end{align}
where $S_{i+}^\alpha$ and $S_{i-}^\alpha$ are symmetric and antisymmetric components of the spin operator
with respect to the left and right sides of a dimer, respectively.
They are defined by
\begin{align}
S_{i\pm}^\alpha = \frac{1}{\sqrt{2}} \left( S_{il}^\alpha \pm S_{ir}^\alpha \right).
\end{align}
The symmetric ($+$) component connects the triplet states,
while the antisymmetric ($-$) component connects the singlet and triplet states.
To resolve the position dependence of the magnetic excitation,
we introduce the following $i_x$-side-dependent dynamical spin correlation function:
\begin{align}
S_{i_x\pm}^{\alpha\beta}(q,\omega) = \frac{1}{2\pi} \int_{-\infty}^\infty \rd t e^{i\omega t}
\langle S_{i_x\pm}^\alpha(q) S_{i_x\pm}^\beta(-q,t) \rangle.
\label{eqn:dynamical-spin}
\end{align}
Since we are interested in the one-magnon process, the spin operator in Eq. (\ref{eqn:Fourier-spin}) is written as
\begin{align}
S^\alpha_{i\pm} \rightarrow
\sum_{m=1}^3 \left[ S_{m0}^\alpha(i_x,\pm) a_{im}^\dagger + S_{0m}^\alpha(i_x,\pm) a_{im} \right],
\label{eqn:one-magnon}
\end{align}
with
\begin{align}
S_{mn}^\alpha(i_x,\pm) = \frac{1}{\sqrt{2}} \left[ S_{mn}^\alpha(i_x,l) \pm S_{mn}^\alpha(i_x,r) \right].
\end{align}
Substituting Eq. (\ref{eqn:one-magnon}) into Eq. (\ref{eqn:Fourier-spin}), we obtain
\begin{align}
S_{i_x\pm}^\alpha(q) =
\sum_{m=1}^3 \left[ S_{m0}^\alpha(i_x,\pm) a_{q i_x m}^\dagger + S_{0m}^\alpha(i_x,\pm) a_{q i_x m} \right].
\label{eqn:Fourier-spin-2}
\end{align}

Next, we replace the $a_q$ boson with the $\alpha_q$ boson using the Bogoliubov transformation.
For this purpose, we write the $l$th ($l=1,\cdots,3N_x$) eigenvector of Eq. (\ref{eqn:eigen}) as
\begin{align}
\vec{X}_{q l} =
\left(
  \begin{array}{c}
    \bu_{q l} \cr
    \bv_{q l}
  \end{array}
\right),
\end{align}
where $\bu_{q l}$ and $\bv_{q l}$ are 3$N_x$-dimensional vectors.
The eigenvectors should be normalized to satisfy the bosonic commutation relation of $\alpha_{q l}$.
For the $l$th eigenvector, the normalization is carried out to satisfy
\begin{align}
\sum_{n=1}^{3N_x} \left( \left| u_{q ln} \right|^2 - \left| v_{q ln} \right|^2 \right) = 1.
\end{align}
From $\vec{X}_{q l}$, we define the following $6N_x\times 6N_x$ marix:
\begin{align}
\hat{X}_q =
\left(
  \begin{array}{cccccc}
    \bu_{q 1} & \cdots & \bu_{q,3N_x} & \bv_{-q 1}^* & \cdots & \bv_{-q,3N_x}^* \cr
    \bv_{q 1} & \cdots & \bv_{q,3N_x} & \bu_{-q 1}^* & \cdots & \bu_{-q,3N_x}^* \cr
  \end{array}
\right).
\end{align}
The $a_q$ boson is then expressed as
\begin{align}
\left(
  \begin{array}{c}
    \ba_q \cr
    \ba_{-q}^\dagger
  \end{array}
\right)
&= \left( \hat{X}_q^{\rm T} \right)^{-1}
\left(
  \begin{array}{c}
    \balpha_q \cr
    \balpha_{-q}^\dagger
  \end{array}
\right) \cr
&\equiv
\left(
  \begin{array}{cc}
    \bU_q & \bV_q \cr
    \bV_{-q}^* & \bU_{-q}^*
  \end{array}
\right)
\left(
  \begin{array}{c}
    \balpha_q \cr
    \balpha_{-q}^\dagger
  \end{array}
\right).
\label{eqn:inverse}
\end{align}
Here, $\balpha_q$ and $\balpha_{-q}^\dagger$ are the 3$N_x$-dimensional vectors
corresponding to the 3$N_x$ positive-energy eigenstates of Eq. (\ref{eqn:eigen}).
Here, $\bU_q$ and $\bV_q$ are $3N_x\times 3N_x$ matrices.

Substituting Eq. (\ref{eqn:inverse}) into Eq. (\ref{eqn:Fourier-spin-2}),
we obtain the following form of the dynamical spin correlation function:
\begin{align}
S_{i_x\pm}^{\alpha\beta}(q,\omega) = \sum_{l=1}^{3N_x}
I^{\alpha\beta}_{i_x\pm}(q,l) \delta \left( \omega - E_{q l} \right).
\end{align}
Here, $E_{q l}$ is the excitation energy of the $l$th mode and $I_{i_x\pm}^{\alpha\beta}(q,l)$ represents its intensity.
The intensity is given by the following form:
\begin{align}
&I_{i_x\pm}^{\alpha\beta}(q,l) \\
&=\sum_{m=1}^3
\left[ S_{0m}^\alpha(i_x,\pm) \left( \bU_q \right)_{m_{i_x} l} 
     + S_{m0}^\alpha(i_x,\pm) \left( \bV_{-q}^* \right)_{m_{i_x} l} \right] \cr
&\times\sum_{n=1}^3
\left[ S_{0n}^\beta(i_x,\pm) \left( \bU_q \right)_{n_{i_x} l} 
     + S_{n0}^\beta(i_x,\pm) \left( \bV_{-q}^* \right)_{n_{i_x} l} \right]^*.
\nonumber
\end{align}
Here, the matrix elements are expressed as $(\cdots)_{m_{i_x} l}$ with $m_{i_x}$ depending on the $i_x$ site.
The explicit form is defined by
\begin{align}
&m_{i_x} = m + 3 ( i_x -1 ).
\end{align}



\begin{thebibliography}{99}

\bibitem{Fujita-1996}
  M. Fujita, K. Wakabayashi, K. Nakada, and K. Kusakabe,
  J. Phys. Soc. Jpn. {\bf 65}, 1920 (1996).

\bibitem{Nakada-1996}
  K. Nakada, M. Fujita, G. Dresselhaus, and M. S. Dresselhaus,
  Phys. Rev. B {\bf 54}, 17954 (1996).




\bibitem{Kobayashi-2005}
  Y. Kobayashi, K. Fukui, T. Enoki, K. Kusakabe, and Y. Kaburagi,
  Phys. Rev. B {\bf 71}, 193406 (2005).

\bibitem{Niimi-2006}
  Y. Niimi, T. Matsui, H. Kambara, K. Tagami, M. Tsukada, and H. Fukuyama,
  Phys. Rev. B {\bf 73}, 085421 (2006).




\bibitem{Ryu-2002}
  S. Ryu and Y. Hatsugai,
  Phys. Rev. Lett. {\bf 89}, 077002 (2002).

\bibitem{Hatsugai-2009}
  Y. Hatsugai,
  Solid State Commun. {\bf 149}, 1061 (2009).

\bibitem{Yao-2009}
  W. Yao, S. A. Yang, and Q. Niu,
  Phys. Rev. Lett. {\bf 102}, 096801 (2009).
  


\bibitem{Delplace-2011}
  P. Delplace, D. Ullmo, and G. Montambaux,
  Phys. Rev. B {\bf 84}, 195452 (2011).



\bibitem{Zhang-2013}
  L. Zhang, J. Ren, J.-S. Wang, and B. Li,
  Phys. Rev. B {\bf 87}, 144101 (2013).

\bibitem{Mook-2014}
  A. Mook, J. Henk, and I. Mertig,
  Phys. Rev. B {\bf 90}, 024412 (2014).

\bibitem{Chisnell-2015}
  R. Chisnell, J. S. Helton, D. E. Freedman, D. K. Singh, R. I. Bewley, D. G. Nocera, and Y. S. Lee,
  Phys. Rev. Lett. {\bf 115}, 147201 (2015).



\bibitem{You-2008}
  J.-S. You, W.-M. Huang, and H.-H. Lin,
  Phys. Rev. B {\bf 78}, 161404 (2008).
  \label{ref:You}




\bibitem{Smirnova-2009}
  O. Smirnova, M. Azuma, N. Kumada, Y. Kusano, M. Matsuda, Y. Shimakawa, T. Takei, Y. Yonesaki, and N. Kinomura,
  J. Am. Chem. Soc. {\bf 131}, 8313 (2009).

\bibitem{Okubo-2010}
  S. Okubo, F. Elmasry, W. Zhang, M. Fujisawa, T. Sakurai, H. Ohta, M. Azuma, O. A. Sumirnova, and N. Kumada,
  J. Phys.: Conf. Ser. {\bf 200}, 022042 (2010).

\bibitem{Onishi-2012}
  N. Onishi, K. Oka, M. Azuma, Y. Shimakawa, Y. Motome, T. Taniguchi, M. Hiraishi, M. Miyazaki, T. Masuda, A. Koda, K. M. Kojima, and R. Kadono,
  Phys. Rev. B {\bf 85}, 184412 (2012).




\bibitem{Kandpal-2011}
  C. H. Kandpal and J. van den Brink,
  Phys. Rev. B {\bf 83}, 140412(R) (2011).

\bibitem{Ganesh-2011}
  R. Ganesh, S. V. Isakov, and A. Paramekanti,
  Phys. Rev. B {\bf 84}, 214412 (2011).

\bibitem{Oitmaa-2012}
  J. Oitmaa and R. R. P. Singh,
  Phys. Rev. B {\bf 85}, 014428 (2012).



\bibitem{Shiina-2003}
  R. Shiina, H. Shiba, P. Thalmeier, A. Takahashi, and O. Sakai,
  J. Phys. Soc. Jpn. {\bf 72}, 1216 (2003).

\bibitem{Sommer-2001}
  T. Sommer, M. Vojta, and K. W. Becker,
  Eur. Phys. J. B {\bf 23}, 329 (2001).
  
\bibitem{Matsumoto-2002}
  M. Matsumoto, B. Normand, T. M. Rice, and M. Sigrist,
  Phys. Rev. Lett. {\bf 89}, 077203 (2002).

\bibitem{Matsumoto-2004}
  M. Matsumoto, B. Normand, T. M. Rice, and M. Sigrist,
  Phys. Rev. B {\bf 69}, 054423 (2004).

\bibitem{Sachdev-1990}
  S. Sachdev and R. N. Bhatt,
  Phys. Rev. B {\bf 41}, 9323 (1990).



\bibitem{hopping-pm}
  The sign of $\bm{B}_k$ is reversed from that for Eq. (\ref{eqn:AB}),
  since the sign of the pair term for the $S^z=\pm$ triplets is opposite to that for the $S^z=0$ triplet
  as shown in Eq. (\ref{eqn:Hk}).



\bibitem{hc}
  Precisely speaking, the lower critical field for a finite system slightly differs from that for the bulk system.
  In the case of $N_x=12$, it is estimated as $h_{\rm c}\simeq 1.02 h_{\rm c(bulk)}$.
  Here, $h_{\rm c(bulk)}=\sqrt{J^2 - 3JJ'}$ is the critical field for the bulk system.

\bibitem{jpc}
  Precisely speaking, the critical interdimer interaction for a finite system slightly differs from that for the bulk system.
  In the case of $N_x=12$, it is estimated as $J'_{\rm c}\simeq 1.02 J'_{\rm c(bulk)}$.
  Here, $J'_{\rm c(bulk)}=\frac{1}{3}J$ is the critical field for the bulk system.



\bibitem{pair}
  Note that the energy of the bulk excitations relative to the potential energy
  is related to the hopping and pair terms in the Hamiltonian.
  For the hopping term, the bulk excitations appear both above and below the potential energy,
  while they appear only below the potential energy for the pair term in bosonic excitations.
  In the case of fermions, such as in superconductors,
  the excitations appear only above the potential energy for the pair term.




\bibitem{Weihong-1991}
  Z. Weihong, J. Oitmaa, and C. J. Hamer,
  Phys. Rev. B {\bf 44}, 11869 (1991).



\end{thebibliography}
\end{document}